\documentclass[aoas,MSNbibl,nameyear,seceqn,dvips]{arximspdf}
\usepackage{multirow}
\usepackage{dcolumn}
\usepackage{stfloats}
\usepackage{graphicx}

\doi{10.1214/14-AOAS744} %kopijuoti is PTS
\volume{8}
\issue{3}
\pubyear{2014}
\firstpage{1341}
\lastpage{1371}

\makeatletter
\newcolumntype{d}[1]{D{.}{.}{#1}}
\fnbelowfloat
\newcommand{\eqref}[1]{(\ref{#1})}
\renewcommand{\mid}{|}

\def\P{\mathbb{P}}
\def\N{\mathrm{N}}
\makeatother

\begin{document}
\begin{frontmatter}

\title{A location-mixture autoregressive model for online forecasting
of lung tumor motion}
\runtitle{Nonlinear online forecasting of lung tumor motion}

\begin{aug}
\author[A]{\fnms{Daniel} \snm{Cervone}\corref{}\ead[label=e1]{dcervone@fas.harvard.edu}\thanksref{M1}},
\author[A]{\fnms{Natesh S.} \snm{Pillai}\ead[label=e2]{pillai@stat.harvard.edu}\thanksref{M1,T2}},
\author[B]{\fnms{Debdeep} \snm{Pati}\ead[label=e3]{debdeep@stat.fsu.edu}\thanksref{M2}},\\
\author[C]{\fnms{Ross} \snm{Berbeco}\ead[label=e4]{rberbeco@lroc.harvard.edu}\thanksref{M3,T4}}
\and
\author[C]{\fnms{John Henry} \snm{Lewis}\ead[label=e5]{jhlewis@lroc.harvard.edu}\thanksref{M3,T5}}
\runauthor{D. Cervone et al.}
\affiliation{Harvard University\thanksmark{M1},
Florida State University\thanksmark{M2},
Brigham and Women's Hospital, Dana-Farber Cancer Institute and Harvard
Medical School\thanksmark{M3}}
\address[A]{D. Cervone\\
N. S. Pillai\\
Department of Statistics\\
Harvard University\\
1 Oxford St.\\
Cambridge, Massachusetts 02138\\
USA\\
\printead{e1}\\
\phantom{E-mail: }\printead*{e2}} %adresu isvedimo komanda gale!
\address[B]{D. Pati\\
Department of Statistics\\
Florida State University\\
117 N Woodward Ave\\
P.O. Box 3064330\\
Tallahassee, Florida 32306\\
USA\\
\printead{e3}}
\address[C]{R. Berbeco\\
J. H. Lewis\\
Radiation Oncology\\
Brigham and Women's Hospital,\\
Dana-Farber Cancer Institute\\
and\\
Harvard Medical School\\
75 Francis St., ASBI-L2\\
Boston, Massachusetts 02115\\
USA\\
\printead{e4}\\
\phantom{E-mail: }\printead*{e5}}
\end{aug}
\thankstext{T2}{Supported in part by the NSF Grant DMS-11-07070.}
\thankstext{T4}{Supported in part by R21CA156068 from the National
Cancer Institute.}
\thankstext{T5}{Supported in part by Award Numbers RSCH1206 from the
Radiological Society of North America.}

% HISTORY:
\received{\smonth{9} \syear{2013}}
\revised{\smonth{4} \syear{2014}}

% ABSTRACT
%
\begin{abstract}
Lung tumor tracking for radiotherapy requires real-time, multiple-step
ahead forecasting of a quasi-periodic time series recording
instantaneous tumor locations. We introduce a location-mixture
autoregressive (LMAR) \mbox{process} that admits multimodal conditional
distributions, fast approximate inference using the EM algorithm and
accurate multiple-step ahead predictive distributions. LMAR outperforms
several commonly used methods in terms of out-of-sample prediction
accuracy using clinical data from lung tumor patients. With its
superior predictive performance and real-time computation, the LMAR
model could be effectively implemented for use in current tumor
tracking systems.
\end{abstract}

% KEYWORDS
\end{frontmatter}

\setcounter{footnote}{3}
%s1 #&#
\section{Introduction}\label{secintro}\label{sec1}
Real-time tumor tracking is a promising recent development in External
Beam Radiotherapy (XRT) for the treatment of lung tumors. In XRT, a~compact linear accelerator is used to deliver photon radiation to the
tumor locations in a narrow beam, minimizing exposure to nearby healthy
tissue. As the location of the lung tumor is in constant motion due to
respiration, some patients who undergo this treatment are implanted
with a small metal marker (known as a fiducial) at the location of a
tumor. During XRT, X-ray imaging reveals the location of the fiducial,
thus providing the desired target of the radiation beam. Tumor tracking
is an advanced technology that minimizes normal tissue exposure by
moving the radiation beam to follow the
tumor position [\citet{rottmann2013markerless,d2005real,schweikard2000robotic}]. However,
there is a system latency of 0.1--1.0 seconds (depending on the
equipment used) that causes the aperture of the radiation beam to lag
behind the real-time location of the tumor. This latency is estimated
empirically by comparing the motion history of the fiducial and
radiation beam aperture.
For tumor tracking XRT to be successful, hardware and software system
latencies must be overcome by the \hyperref[sec1]{Introduction} of a predictive algorithm.

As accurate radiotherapy is essential for both minimizing radiation
exposure to healthy tissue and ensuring the tumor itself is
sufficiently irradiated, the subject of predicting tumor motion to
overcome the system latency has received a good deal of attention in
the medical community. Any possible forecasting approach must provide
\mbox{$k$-}step ahead predictive distributions in real time, where $k$ is
approximately equal to the system latency multiplied by the sampling
frequency of the tumor tracking imagery. Real-time forecasting requires
that a (\mbox{$k$-}step ahead) prediction be made before any further data on
the tumor's motion has been recorded.

Statistical methods for tumor prediction in the literature include
penalized linear models [e.g., \citet{sharpprediction2004} and
many others], the Kalman filter [\citet{murphy2002adaptive}],
state--space models [\citet{kaletstate-based2010}] and wavelets
[\citet{ernst2007prediction}]; machine learning methods include
kernel density estimation [\citet{ruanonline2010}], support vector
regression [\citet{riazpredicting2009,ernstforecasting2009}] and
neural networks [\citet
{murphy2002adaptive,murphycomparative2006}]. All of these examples
include simulations of out-of-sample prediction using real patient data
in order to assess forecasting accuracy. Because predictive performance
varies considerably from patient to patient and across different
equipment configurations, of particular importance to the literature
are comparisons of different prediction methods for the same set of
patients with the same conditions for data preprocessing [\citet
{sharpprediction2004,krausscomparative2011,ernst2013evaluating}]. While
standard ``off-the-shelf'' time series forecasting models can be
applied to lung tumor tracking, better predictive performance can be
achieved with a model that explicitly incorporates the dynamics of
respiratory motion.

We propose a novel time series model which we call a location-mixture
autoregressive process (LMAR). A future observation ($Y_n$) given the
observed history of the time series is assumed to follow a Gaussian mixture,
%
%e1.1 #&#
\begin{equation}
\label{LMARintro} Y_n | Y_{n-1}, Y_{n-2}, \ldots\sim
\sum_{j=1}^{d_n} \alpha_{n,j} \N\bigl(
\mu_{n,j}, \sigma^2\bigr),
\end{equation}
where $\sum_{j=1}^{d_n} \alpha_{n,j} = 1$ and $\mu_{n,j}$ is of the form
%
%e1.2 #&#
\begin{equation}
\label{mixturemeans} \mu_{n,j} = \tilde{\mu}_{n,j} + \sum
_{l=1}^p \gamma_l Y_{n-l}.
\end{equation}
We refer to this as a location-mixture autoregressive model because the
autoregressive part of the component means, $\sum_{l=1}^p \gamma_l
Y_{n-l}$, is the same for all $j$, and only the location parameter,
$\tilde{\mu}_{n,j}$, changes across the components in \eqref
{LMARintro}. Our model differs from other time series models that yield
mixture-normal conditional distributions (e.g., the class of threshold
autoregressive models [\citet{tongthreshold1980}], \mbox{including}
Markov-switching autoregressive models [\citet{hamiltonnew1989}]
and the mixture autoregressive models of \citet{wongmixture2000})
in that $\tilde{\mu}_{n,j}$ in~\eqref{mixturemeans} depends on an
unknown subseries of the time series, at least $p$ observations in the
past. The mixture weights, $\{\alpha_{n,j}\}$, also depend on the
entire history of the observed time series, and the number of mixture
components in our model, $d_n$, increases with $n$.

Another noteworthy characteristic of our model is that all parameters
in \eqref{LMARintro} are obtained from a single, unknown $(p + 1)
\times(p+1)$ positive definite matrix. This parsimonious
parameterization is motivated in part by the need for real-time
parameter estimation and forecasting. Compared with other mixture
autoregressive models, LMAR is simpler to fit and admits accurate
closed-form expressions for \mbox{$k$-}step ahead predictive distributions.
While the data application we consider shows the promise and appeal of
the LMAR model, we believe a thorough treatment of its theoretical
properties (a future endeavor) is necessary before the LMAR model is a
viable ``off-the-shelf'' method for diverse data sets.

We motivate our model in the context of time series motifs, which offer
a geometric interpretation of the components in our model. In general
terms, motifs catalog recurring patterns in time series and are
commonly used in data mining tasks for which a symbolic representation
of a time series is useful, such as event detection and time series
clustering or classification [\citet
{lonardifinding2002,ye2009time,tanaka2005discovery,fu2011review}]. For
the purposes of forecasting, predictive state representations
[\citet{littman2002predictive,shalizi2003optimal,boots2011online}]
categorize time series motifs not as subseries of the observed data,
but as equivalence classes of conditional predictive distributions.

%Time series motifs can also be used for forecasting, though the
%literature is sparse on this subject. The LMAR model is unique in that
%motifs are instantiated within a well-defined data-generating process,
%as opposed to being algorithmically extracted from the data.

Section~\ref{secdata} of this paper discusses the important features of
the data we use and graphically motivates our model. Section~\ref{secLMAR} formally introduces the LMAR model and describes parameter
estimation and forecasting using principled methods that are feasible
in real time. Section~\ref{secevaluation} describes the procedure for
comparing out-of-sample prediction error under our model with competing
forecasting methods for tumor tracking, including the selection of
tuning parameters. The results of this comparison are discussed in
Section~\ref{secresults}, and Section~\ref{secdiscussion} summarizes
and points out future directions.

%s2 #&#
\section{Tumor tracking data}\label{secdata}

We have data on 11 patients treated at the Radiation Oncology Clinic at
the Nippon Telegraph and Telephone Corporation Hospital in Sapporo,
Japan. A detailed discussion of the conditions and instruments involved
in the data acquisition is available in \citet
{berbecoresidual2005}. The data is derived from observations of the
position of gold fiducial markers implanted into the tumors of lung
cancer patients. The marker position is determined via stereoscopic
X-ray imaging conducted at 30 Hz. In each of the two stereoscopic
images, the marker position is automatically detected using
thresholding and edge detection. The position of the marker in these
two images is used to triangulate its position in 3D space relative to
the radiation beam. Data consists of tumor positions measured over one
or multiple days of radiotherapy treatment delivery (range 1--12), and
for multiple sequences on each day, denoted \textit{beams}. In our data
set, there are a total of 171 such distinct sequences, with lengths
varying from 637 observations (about 21 seconds at 30 observations per
second) to 8935 observations (about 5 minutes).

 %Our data set is
%provided in this paper's supplementary materials [\citet{cervone-data}].
%Table~\ref{tabsummary} provides summary statistics for each patient's
%data.

Note that this paper focuses on within-beam forecasting---that is, each
beam is treated independently and there is no information sharing
between patients or within different beams from the same patient.
Developing methodology for combining prediction models from distinct
time series (both within and across patients) is an important area for
further research.

%s2.1 #&#
\subsection{Features of the data}\label{subsecfeatures}
\begin{figure}

\includegraphics{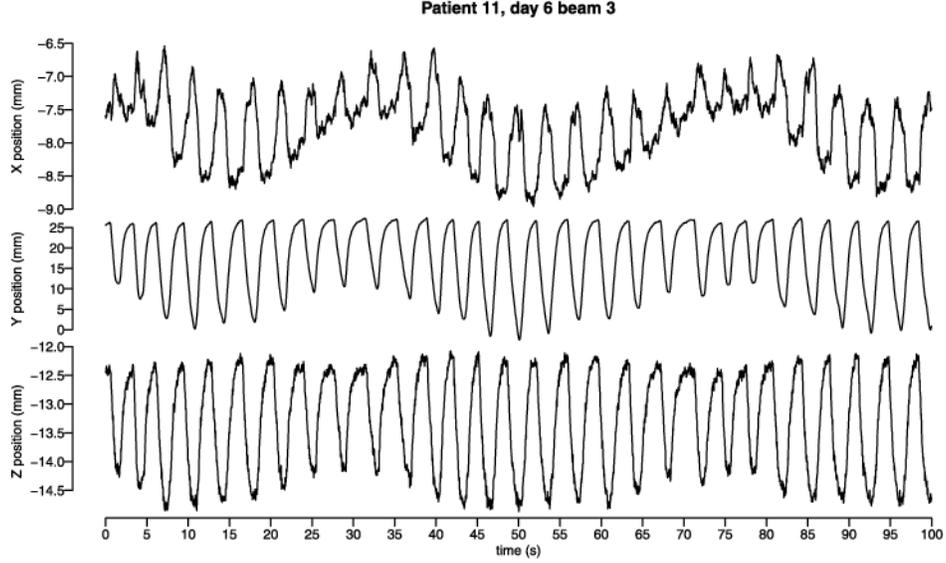}

\caption{Sample time series of 3D locations of lung tumor. The $X$ axis
is the lateral--medial \mbox{(left--right)} direction, $Y$ axis
superior--inferior, and $Z$ axis anterior--posterior.}\label{xyzplot}
\end{figure}

%In this data set, the 3-dimensional lung tumor coordinates are as
%follows. X is the lateral--medial (left--right) direction, Y is the
%superior--inferior direction and Z is the anterior--posterior direction.
Each observation in each sequence is a point in~$\mathbb{R}^3$,
representing the real-time 3D location of the lung tumor. The $X$ axis
is the lateral--medial (left--right) direction, the $Y$ axis is
superior--inferior, and the $Z$ axis is anterior--posterior, with all
measurements in millimeters.\footnote{The origin is set to the
isocenter, which is the center of rotation for the linear accelerator
axis motions. During treatment, the patient is positioned so that this
coincides with the centroid of the region being treated. However, there
is uncertainty in determining this point, so the data is best thought
of as relative tumor motion on each day.} Figure~\ref{xyzplot} shows
the motion in each dimension during the first 100 seconds of a
particular observation sequence. As expected with respiratory motion,
the pattern is approximately periodic, with inhalation closely
corresponding to decreasing values in the $Y$ direction. However, the
amplitude of each breath varies considerably (in Figure~\ref{xyzplot}
the variation seems periodic, though this is not a typical feature of
the data). The curves undergo gradual baseline location shifts and,
while it may not be visually discerned from Figure~\ref{xyzplot}, it is
common for respiratory cycles to change periodicity, either
sporadically or gradually over time. Table~\ref{tabsummary} shows the
variability in period and amplitude of the respiratory traces, both
within and between patients.

%
%f1 #&#

%
%t1 #&#
\begin{table}[b]
\tabcolsep=15pt
\caption{Summary statistics for the first principal component of respiratory trace data, at the patient level}\label{tabsummary}
\begin{tabular*}{\tablewidth}{@{\extracolsep{\fill}}@{}ld{2.0}d{4.2}d{2.2}ccc@{}}
\hline
& & & \multicolumn{2}{c}{\textbf{Amplitude (mm)}} & \multicolumn{2}{c@{}}{\textbf{Period(s)}}\\[-6pt]
 & \multicolumn{1}{c}{\multirow{2}{24pt}{\centering{\textbf{Total beams}}}} & \multicolumn{1}{c}{\multirow{2}{29pt}{\centering{\textbf{Total time (s)}}}} & \multicolumn{2}{c}{\hrulefill} & \multicolumn{2}{c@{}}{\hrulefill}\\
\textbf{Patient} &  &  & \multicolumn{1}{c}{\textbf{Mean}} & \multicolumn{1}{c}{\textbf{SD}} & \multicolumn{1}{c}{\textbf{Mean}} & \textbf{SD}\\
\hline
\phantom{0}1 & 4 & 212.27 & 14.57 & 6.98 & 3.66 & 1.16 \\
\phantom{0}2 & 2 & 136.87 & 13.74 & 1.84 & 3.89 & 1.06 \\
\phantom{0}3 & 2 & 80.93 & 9.84 & 3.16 & 3.97 & 0.56 \\
\phantom{0}4 & 38 & 2502.67 & 8.86 & 1.35 & 2.88 & 0.31 \\
\phantom{0}5 & 26 & 2769.33 & 7.90 & 1.66 & 3.61 & 0.68 \\
\phantom{0}6 & 28 & 2471.93 & 10.07 & 2.51 & 2.58 & 0.55 \\
\phantom{0}7 & 11 & 1661.37 & 9.66 & 2.41 & 5.05 & 1.09 \\
\phantom{0}8 & 8 & 832.80 & 14.38 & 4.02 & 3.15 & 1.18 \\
\phantom{0}9 & 15 & 2599.90 & 11.45 & 1.61 & 3.09 & 0.41 \\
10 & 15 & 3497.67 & 14.88 & 3.65 & 3.77 & 0.64 \\
11 & 22 & 3674.77 & 21.81 & 5.05 & 3.38 & 0.52 \\
\hline
\end{tabular*}
\end{table}

\begin{figure}%[h!]

\includegraphics{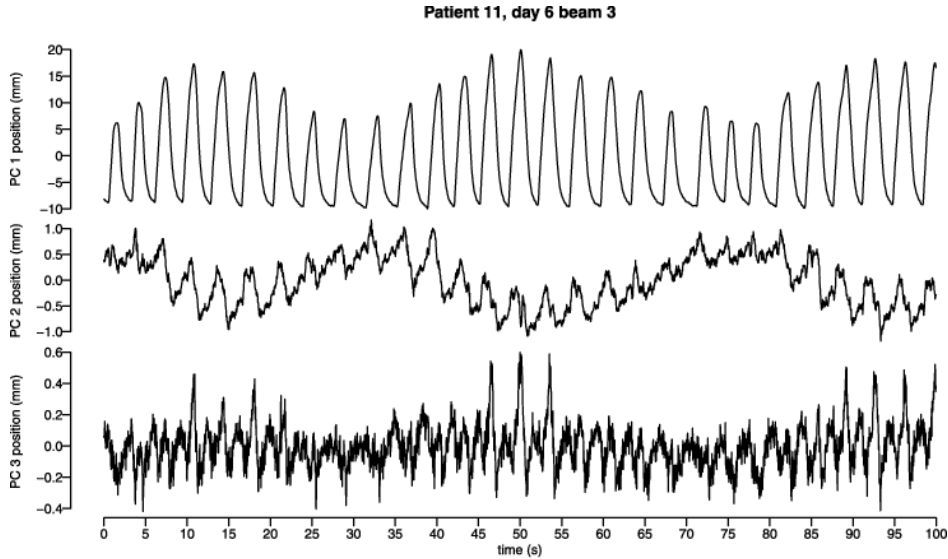}

\caption{Time series of principal components. Components 2 and 3
exhibit periodic behavior, but with much smaller magnitude.}
\label{PCplot}
\end{figure}

Due to the extremely high correlations between series of observations
from different dimensions, it is useful to consider a lower-dimensional
representation of the 3D process. Transforming each 3D sequence into
orthogonal components using principal component analysis (PCA) loads
the periodic respiratory dynamics onto the first component,
representing about $99\%$ of the total variance in the 3D data. The
last two principal components still exhibit some periodic behavior (see
Figure~\ref{PCplot}), but the signal is weak relative to the
noise.\footnote{A referee pointed out that while the first principal
component gives the linear combination of the 3D data with maximum
variance, it is not necessarily the most \textit{forecastable} linear
combination. Alternative linear transformations (e.g., forecastable
components [\citet{goerg2013forecastable}]) may load additional
periodic features to the first component than we observe with PCA. In
choosing an appropriate transformation, the goal is to find an
orthogonal basis in which componentwise predictions have the smallest
error when transformed back to the original basis. We do not explore
this issue here; however, one advantage in using the first principal
component is that the signal-to-noise ratio will be high, allowing for
forecast procedures that are not well suited for measurement error in
the observed data.} In addition to dimension reduction and useful
interpretability, the PCA transformation prevents any loss of
statistical efficiency if models are fit independently for each
component. \citet{ruanonline2010} compared independent-component
prediction before and after PCA using kernel density estimation,
finding smaller 3D root mean squared prediction error when using the
PCA-transformed data for prediction. When comparing several algorithms
for predicting lung tumor motion, both \citet{ernst2013evaluating}
and \citet{krausscomparative2011} used the principal components,
then transformed their predictions to the original linear basis of the data.

%
%f2 #&#

For the remainder of this study, we focus on modeling the first
principal component only, as it encodes such a large portion of the
system dynamics. In clinical implementation, we would forecast
independently on each orthogonal component and transform back to the
original linear basis in order to inform the location of the radiation
treatment beam.

%s2.2 #&#
\subsection{Time series motifs for forecasting: A graphical example}\label{subsecmotifs}

%JHL: I am not sure if it is useful, but the extended length of exhale
%is common to most breathing patterns, and has is the reason many
%people has used a cos^(2n) model for respiration. (Lujan 1999, A
%method for incorporating organ motion due to breathing into 3D dose
%calculations Med Phys).

\begin{figure}%[b!]

\includegraphics{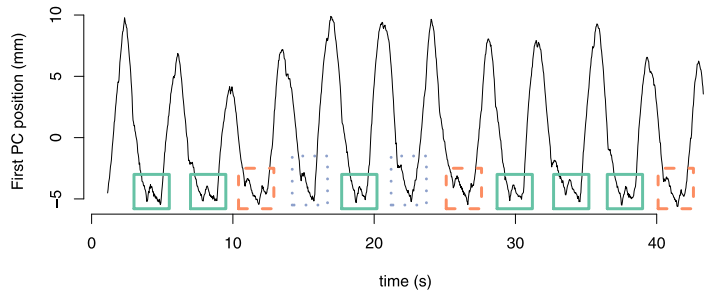}

\caption{Recurring patterns (coded by color and line type) in the
first principal component of patient~10, day 1, beam 3. Areas boxed by
lines of the same color/line type resemble one another. The behavior
highlighted in these motifs is most likely caused by the patient's heartbeat.}
\label{micromotifs}
\end{figure}

Because the data are quasi-periodic, it is useful to look at short
patterns that recur at possibly irregular intervals, which we call
motifs (we provide a more rigorous definition of time series motifs in
Section~\ref{subsecTAR}). Figure~\ref{micromotifs} highlights different
motifs in the first principal component at the end of the exhale (start
of the inhale) for a particular observation sequence. The highlighted
areas appear to be heartbeats, which affect the location of the tumor
differently depending on the real-time location of the tumor relative
to the heart.
%DC: Not totally sure if those are heartbeats
%
%f3 #&#

\begin{figure}

\includegraphics{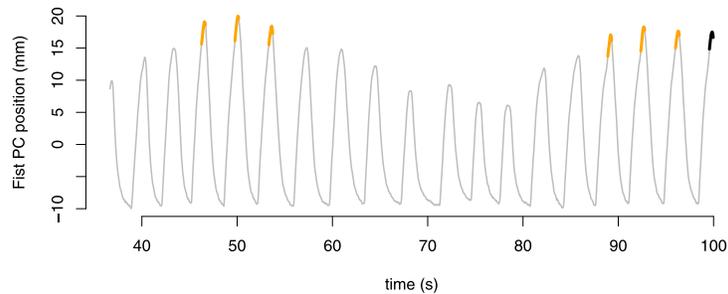}

\caption{The most recent 0.43~s (13 observations) are in black. The
thicker orange segments share similar local history.}
\label{predictexample1}
\end{figure}

Observing repeated patterns within each time series in the data
suggests a modeling/prediction framework that leverages this structure.
In general, if the recent past of the time series resembles a motif we
have observed previously in the data, then the shape of this motif
should inform our predictions of future observations; this idea is
formalized through predictive state representations [\citet
{littman2002predictive,shalizi2003optimal}]. For a graphical
illustration, consider predicting 0.4~s (12 steps) ahead for the first
principal component of the curve displayed in Figure~\ref{PCplot}. We
have observed 100 seconds of the process, and it appears as though we
have just observed the start of the exhale; the current observation at
time $t=100$ seconds, as well as the previous 12 observations, are
colored orange in Figure~\ref{predictexample1}. Colored in black are
segments earlier in the time series that resemble the current motif
(specifically, we highlighted subseries of length 13 where the tenth
point has the largest magnitude, and the 11th--13th points are decreasing).
%
%f4 #&#

%
%f5 #&#
\begin{figure}[b]

\includegraphics{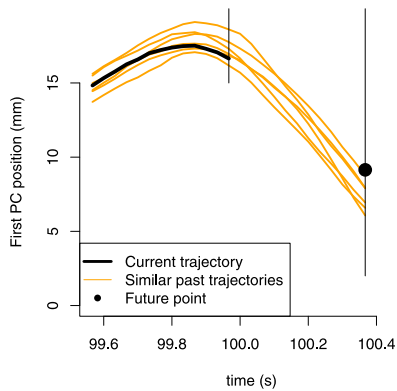}

\caption{The recent history of the process (thick black line)
instantiates a motif. Previous instances of this motif, and their
subsequent evolutions, are in orange and provide reasonable predictions
for future points (black dot).}
\label{predictexample2}
\end{figure}

To predict future observations, we can incorporate the points
immediately succeeding the
endpoints of black motifs. Figure~\ref{predictexample2} shows these
trajectories (in gray), and the actual current trajectory of the
process is shown in orange, with a point giving the value 0.4~s in the
future. The gray curves provide reasonable forecasts for the future
evolution of the time series and, indeed, the actual future value is
close to where these trajectories predict.

Our model, formally introduced in Section~\ref{secLMAR}, implements the
forecasting approach sketched in this subsection using an
autoregressive model for the data-generating process.

\section{Location-mixture autoregressive processes}\label{secLMAR}

Here, we define the LMAR process and provide computationally efficient
algorithms for parameter estimation and \mbox{$k$-}step ahead forecasting. To
establish terminology, we denote a \textit{time series} as an ordered
sequence of real numbers $\{Y_i \in\mathbb{R}, i = 0, \pm1, \pm2,
\ldots\}$ measured at regular, equally spaced intervals. Also, a
\textit{subseries} of length $p+1$ is a subset of a time series $\{Y_i,
i=0, \pm1, \ldots\}$ comprised of consecutive observations, $Y_{i},
Y_{i+1}, \ldots, Y_{i+p}$. For notational ease, we will denote the
subseries as $Y_{i\dvtx(i+p)}$ or, equivalently, $Y_{i+0\dvtx p}$.

%s3.1 #&#
\subsection{A model for the data-generating process}\label{subsecDGP}

Let $\{Y_i, i=-m,\ldots,n\}$ be a time series. Also, assume $\Sigma$ is
a $(p+1) \times(p+1)$ symmetric, nonnegative definite matrix, where
$\Sigma_{11}$ is the upper-left $p \times p$ submatrix, $\Sigma_{22}$
is the single bottom-right element, and $\Sigma_{21}$ and $\Sigma_{12}$
are the respective off-diagonal row and column vectors. $p$ is assumed
to be fixed and known. For notational ease, let $\gamma= \Sigma
_{11}^{-1}\Sigma_{12}$, $\sigma^2 = \Sigma_{22} - \gamma'\Sigma_{12}$,
and $\mathcal{J}_i = \{p+1, \ldots, i+m-p\}$. Last, let
\[
V_{ij} =
\pmatrix{ Y_{i-p} -
Y_{i-j-p}
\vspace*{3pt}\cr
\vdots
\vspace*{3pt}\cr
Y_{i-2} - Y_{i-j-2}
\cr
Y_{i-1} - Y_{i-j-1}}.
\]
As in \eqref{LMARintro}, we assume that the distribution of $Y_i$ given
$Y_{-m}, \ldots, Y_{i-1}$ is a normal mixture,
%
%e3.1 #&#
%e3.2 #&#
\begin{eqnarray}
\label{DGP} Y_i | Y_{(-m)\dvtx(i-1)} &\sim&\sum
_{j \in\mathcal{J}_i} \alpha_{i,j} \mathrm{N}\bigl(
\mu_{i,j}, \sigma^2\bigr)
\nonumber\\[-8pt]\\[-8pt]
\eqntext{\displaystyle\mbox{where } \alpha_{i,j}  = \frac{ \exp( -({1}/{2})V_{ij}'\Sigma
_{11}^{-1}V_{ij} ) } {
\sum_{l \in\mathcal{J}_i} \exp (
-({1}/{2})V_{il}'\Sigma_{11}^{-1}V_{il}
) }
\mbox{ and }\mu_{i,j}  = Y_{i-j} + \gamma'V_{ij}.}
\end{eqnarray}

The model in \eqref{DGP} defines the location-mixture autogressive
process with parameter $\Sigma$ [abbreviated LMAR($\Sigma$)]. We can
recognize the location-mixture form originally given in \eqref
{LMARintro} by writing $\mu_{i,j} = \tilde{\mu}_{i,j} + \sum_{l=1}^p
\gamma_l Y_{i-l}$, where
%
%e3.3 #&#
\begin{equation}
\label{locationparam} \tilde{\mu}_{i,j} = Y_{i-j} - \sum
_{l=1}^p \gamma_l Y_{j-l}
\end{equation}
and $(\gamma_p \gamma_{p-1} \cdots \gamma_1)' = \gamma$. Thus, the
distribution for $Y_i|Y_{(-m)\dvtx(i-1)}$ is a normal mixture with
$|\mathcal{J}_i|$ different mean components---each sharing a common
autoregressive component but different location parameter---equal
variance across components ($\sigma^2$) and data-driven mixture weights
($\alpha_{i,j}$). We assume \eqref{DGP} for all $i \geq0$, but we do
not make any distributional assumptions about $Y_{(-m)\dvtx(-1)}$.

%It may also be illustrative to write the distribution function of
%$Y_i|Y_{(-m)\dvtx(i-1)}$\dvtx
% & F(y|Y_{(-m)\dvtx(i-1)}) =
% \\
% \sum_{j \in\mathcal{J}_i} &
% \Phi\left(
% \frac{ y - (Y_{i-j} + \gamma' V_{ij}) } { \sigma}
% \right)
% \left(
% \frac{ \exp\left( -\frac{1}{2}V_{ij}'\Sigma_{11}^{-1}V_{ij} \right) }
% { \sum_{l \in\mathcal{J}_i} \exp\left(
% -\frac{1}{2}V_{il}'\Sigma_{11}^{-1}V_{il}
% \right) }
% \right),

As $\Sigma$ parameterizes the entire mixture distribution, the
component means and mixture weights are linked through a common
parameter which encourages self-similarity in the data-generating
process. If two subseries $Y_{(i-p)\dvtx(i-1)}$ and $Y_{(i-p-j)\dvtx
(i-1-j)}$ resemble one another in that $V_{ij}'\Sigma_{11}^{-1}V_{ij}$
is small, then we have a large weight on the mixture component with
mean $Y_{i-j} + \gamma'V_{ij}$. This means that the next observation of
the process, $Y_i$, is centered near a previous value of the series
$Y_{i-j}$ inasmuch as the subseries of observations preceding $Y_i$ and
$Y_{i-j}$ have a similar shape. Simply put, if $Y_i$ and $Y_{i-j}$ are
preceded by similar values, then the components of $V_{ij}$ will be
close to 0. This drives up the mixture weight $\alpha_{i,j}$, implying
the mean of $Y_i$ will be close to $\mu_{i,j}$ (which itself is close
to $Y_{i-j}$).

The dimension of $\Sigma$, $p+1$, can in principle be chosen using
standard model selection methods (e.g., Bayes factors), though if the
goal of fitting a LMAR model is prediction, we recommend
cross-validation or hold-out testing for choosing $p$. For
quasi-periodic time series, a reasonable choice for $p$ might be
anywhere between one-tenth and one-third of the average number of
observations per period. Larger values of $p$ increase the
computational load in estimating $\Sigma$ while favoring sparser
component weights.

The model \eqref{DGP} specifies the role of time series motifs in the
data-generating process, which was informally discussed in Section~\ref{subsecmotifs}. To illustrate this, we introduce a latent variable
$M_i$ that takes values in $\mathcal{J}_i$, such that for all $j \in
\mathcal{J}_i$,
%
%e3.4 #&#
\begin{equation}
\label{Mn} \mathbb{P}(M_i = j| Y_{(-m)\dvtx(i-1)}) \propto\exp
\bigl( -\tfrac{1}{2} V_{ij}' \Sigma_{11}^{-1}
V_{ij} \bigr).
\end{equation}
Then, given $M_i=j$, we induce the same distribution for $Y_i$ as in
\eqref{DGP} by assuming
%
%e3.5 #&#
\begin{equation}
\label{fullconditional} Y_i \mid[M_i=j, Y_{(-m)\dvtx(i-1)}] \sim
\N\bigl( Y_{i-j} + \gamma' V_{ij},
\sigma^2 \bigr).
\end{equation}
Expression \eqref{fullconditional} can be used to define a motif
relation: each subseries of length ($p + 1$) is a \textit{motif}, and
$Y_{(i-p)\dvtx i}$ is an \textit{instance} of motif $Y_{(i-p-j)\dvtx(i
- j)}$ if $M_i = j$ [thus yielding \eqref{fullconditional}]. We denote
this by writing
\[
\mbox{(motif) } Y_{(i-p-j)\dvtx(i-j)} \rightarrow Y_{(i-p)\dvtx i}
\mbox{
(instance)}.
\]
Note that our indexing set $\mathcal{J}_i$ is defined in such a way
that instances of a particular motif cannot overlap (share a common
component $Y_j$) with the motif itself.

Our definition of motifs is atypical of the literature for data mining
tasks [\citet{lonardifinding2002}] and predictive state
representations of time series [\citet{littman2002predictive}].
For instance, the relationship that instantiates motifs \mbox{(notated
$\rightarrow$)} is not symmetric and is not an equivalence relation; for
this reason we have defined a motif instance distinctly from a motif.
Also, we define motifs as observed subseries of the data and motif
instances as latent states (we do not observe $M_i$). For most data
mining tasks, time series motifs represent an equivalence class of
observed subseries of the data (possibly transformed) [\citet
{fu2011review}], whereas predictive state representations of time
series treat motifs as latent equivalence classes of predictive
distributions [\citet{shalizi2003optimal}].

However, our definition of motifs preserves the interpretation of
geometric similarity we sketched in Section \ref{subsecmotifs}. From
\eqref{Mn}, we have $M_i = j$ (meaning $Y_{(i-p-j)\dvtx(i-j)}
\rightarrow Y_{(i-p)\dvtx i}$) with\vspace*{1pt} high probability if $V_{ij}$ is
small with respect to the $\Sigma_{11}^{-1}$ inner product norm. Our
model thus expects a subseries that is an instance of a particular
motif to be close to the motif, and $\Sigma$ parameterizes this
distance metric.

%s3.2 #&#
\subsection{Comparison with other mixture autoregressive processes}\label{subsecTAR}

We may compare the LMAR($\Sigma$) to a general form of regime-switching
autoregressive models, for which we can write the distribution function
of $Y_i$ conditional on all available history of the process
$Y_{(-m)\dvtx(i-1)}$ as
%
%e3.6 #&#
\begin{equation}
\label{generalTAR} F(y|Y_{(-m)\dvtx(i-1)}) = \sum_{j=1}^d
\alpha_{i,j} \Phi\biggl( \frac{ y - (\beta_{0,j} + \sum_{l=1}^{p} \beta
_{l,j}Y_{i-l}) } { \sigma_j } \biggr),
\end{equation}
where $\sum_{j=1}^d \alpha_{i,j} = 1$ for all $i$ and $\Phi$ denotes
the standard normal CDF. Models satisfying \eqref{generalTAR} can be
represented in the framework of threshold autoregressive models
[\citet{tong1978threshold,tongthreshold1980}; see \citet
{tongnon-linear1990} for a book-length treatment], which represent
\eqref{generalTAR} using an indicator series $\{M_i\}$ taking values on
$\{1,\ldots,d\}$, such that
%
%e3.7 #&#
\begin{equation}
\label{generalTAR2} Y_i = \beta_{0,M_i} + \sum
_{l=1}^p \beta_{l, M_i}Y_{i-l} +
\sigma_{M_i}\varepsilon_i,
\end{equation}
where $\{\varepsilon_i\}$ are i.i.d. standard normals. Generally, $M$
is not observed, although there are notable exceptions such as the
self-exciting threshold AR model of \citet{tongthreshold1980}.

A canonical model of this form is the mixture autoregressive model of
\citet{lemodeling1996} and \citet{wongmixture2000}, which
assumes $\{M_i\}$ are i.i.d. and independent of $Y$. Another special
case of \eqref{generalTAR2} is when $M$ is a Markov chain, such as in
the Markov-switching autoregressive models of \citet
{hamiltonnew1989} and \citet{mcculloch1994statistical}. More
general stochastic structure for $M$ is considered by \citet
{laubayesian2008}, as well as in mixture-of-experts models in the
machine learning literature [\citet
{carvalhomixtures--experts2005}]. These models seem favorable over the
mixture autoregressive models of \citet{wongmixture2000} when the
data is seasonal or quasi-periodic, as is the case with the time series
we consider.

The LMAR($\Sigma$) process differs from \eqref{generalTAR} in that the
mixture means, following \eqref{DGP}--\eqref{locationparam}, are given by
\begin{eqnarray*}
\mu_{i,j} &=& \tilde{\mu}_{i,j} + \sum
_{l=1}^p \gamma_l Y_{i-l}
= Y_{i-j} + \sum_{l=1}^p
\gamma_l Y_{i-l} - \sum_{l=1}^p
\gamma_l Y_{j-l},
\end{eqnarray*}
instead of $\mu_{i,j} = \beta_{0, j} + \sum_{l=1}^p \beta_{l,
j}Y_{i-l}$ as in \eqref{generalTAR}. Thus, for LMAR($\Sigma$), the
autoregressive coefficients ($\gamma$) are fixed, and the
normal-mixture form of the conditional distribution is induced by a
location shift that is a function of a random subseries of past
observations, $\tilde{\mu}_{i,j}$. The normal-mixture form of \eqref
{generalTAR}, however, is induced by a mixture distribution for
autoregressive coefficients of the same lagged values of the time
series. The mixture weights of the LMAR($\Sigma$) process are also
strongly data driven, depending on the entire history of the process.
Unlike many forms of mixture autoregressive models, there is no prior
distribution or conditional dependence structure assumed for $M$; the
distribution of $M$ is supplied entirely by the data.

Another key difference is that LMAR($\Sigma$) does not assume a fixed
number of mixture components, as is clear from \eqref{DGP}. But because
the same autoregressive coefficient vector ($\gamma$) parameterizes all
mean components $\mu_{i,j}$, we actually have a much smaller parameter
space than all the instances of \eqref{generalTAR} cited above, which
include the parameters for the mixture components ($d$ vectors of
length $p+1$ for the means) as well as for the distribution of $M$. A
small parameter space is advantageous in the context of our data
application, as it facilitates rapid updating. Also, time constraints
will not allow for any goodness-of-fit or model selection procedures
for choosing structural parameters such as $d$ or $p$ in \eqref
{generalTAR}, or structural parameters for $M$. The only structural
parameter in the LMAR($\Sigma$) model is $p$, and in our analysis of
this data set we found that predictive distributions were quite stable
for different choices of $p$.

The most important distinction of the LMAR($\Sigma$) model is the
existence of good approximations for \mbox{$k$-}step ahead predictive
distributions, for $k \leq p$, which are given in Section~\ref{subsecprediction}. Closed-form predictive distributions for $k>1$ are
not available for many models of the form \eqref{generalTAR} [the
exception is the Markov-switching autoregressive models of \citet
{hamiltonnew1989}; for a discussion see \citet
{krolzig2000predicting}]. \citet{wongmixture2000} recommended
Monte Carlo estimates of \mbox{$k$-}step ahead predictive distributions,
although \citet{boshnakov2009analytic} found for them a
closed-form representation as a normal mixture. Calculating the mixture
component parameters for moderate $k$, however, is quite laborious. For
the general model \eqref{generalTAR}, \citet{degooijerrecent1992}
discussed the difficulty in \mbox{$k$-}step ahead forecasting and questioned
whether predictive performance is improved over classes of linear time
series models [also see \citet{tongmulti-step1988} for a
discussion of the robustness of medium-to-long range forecasts using
threshold autoregressive models].

%s3.3 #&#
\subsection{Parameter estimation}\label{subsecestimation}
In order to be able to adjust radiotherapy treatments in real time to
the patient's breathing pattern, we seek estimation procedures that are
fast enough to run online (in less than a few seconds). As a general
rule, this favors approximate closed-form solutions to estimating
equations over exact numerical or Monte Carlo methods. To estimate
$\Sigma$, which is the only unknown parameter of this model, we take a
conditional likelihood approach based on the conditional distribution
$Y_{0\dvtx n}|Y_{(-m)\dvtx(-1)}$. We assume the full-data likelihood
can be written as
\[
L(\psi, \Sigma) = L_1(\psi,
\Sigma)L_2(\Sigma),
\]
where $L_1(\psi, \Sigma) \propto\mathbb{P}(Y_{(-m)\dvtx(-1)}; \psi,
\Sigma)$ and $L_2(\Sigma) \propto\mathbb{P}(Y_{0\dvtx n}|Y_{(-m)\dvtx
(-1)}; \Sigma)$.\break The distribution of the first $m$ observations, and
thus $L_1$, is left unspecified, and all information for $\Sigma$ comes
from $L_2$. If $L_1$ depends on $\Sigma$, there will be some loss of
efficiency when using only $L_2$ for inference versus the complete-data
likelihood, though under mild conditions the maximum conditional
likelihood estimate is consistent and asymptotically efficient
[\citet{kalbfleisch1970application}].

The conditional likelihood, $L_2(\Sigma)$, can be written as
%
%e3.8 #&#
\begin{eqnarray}
\label{conditionallikelihood} L_2(\Sigma) &=& \prod_{i=0}^n
\frac{1}{\sigma} \biggl[ \sum_{j \in\mathcal{J}_i} \exp\biggl( -
\frac{1}{2\sigma^2} \bigl(Y_i - Y_{i-j} -
\gamma'V_{ij}\bigr)^2 \biggr)
\nonumber\\[-8pt]\\[-8pt]
&&\hspace*{54pt}{} \times\biggl( \frac{\exp(-V_{ij}'\Sigma_{11}^{-1}V_{ij}/2)}{\sum_{l
\in\mathcal{J}_i} \exp(-V_{il}'\Sigma_{11}^{-1}V_{il}/2)} \biggr)
\biggr].\nonumber
\end{eqnarray}
To maximize \eqref{conditionallikelihood}, we augment the data to $\{
Y_{0\dvtx n},M_{0\dvtx n}\}$, with $M_i$ as in \eqref{Mn}. This invites
the use of the Expectation--Maximization (EM) algorithm [\citet
{dempstermaximum1977}] to estimate $\Sigma$. The augmented-data
(complete-data) conditional likelihood is
\begin{eqnarray*}
L_{2, \mathrm{com}}( \Sigma) &=& \prod
_{i=0}^n \frac{1}{\sigma} \prod
_{j \in\mathcal{J}_i} \biggl[ \exp\biggl( -\frac{1}{2\sigma^2}
\bigl(Y_i - Y_{i-j} - \gamma'V_{ij}
\bigr)^2 \biggr)
\\
&&\hspace*{54pt} {}\times\biggl( \frac{\exp(-V_{ij}'\Sigma_{11}^{-1}V_{ij}/2)}{\sum_{l
\in\mathcal{J}_i} \exp(-V_{il}'\Sigma_{11}^{-1}V_{il}/2)} \biggr)
\biggr]^{\mathbf{1}[M_i = j]}.
\end{eqnarray*}
This can be simplified further. Let $W_{ij}' = (V_{ij}' Y_i -
Y_{i-j})$, and recalling the notation for $\sigma$ and $\gamma$, we have
%
%e3.9 #&#
\begin{equation}
\label{augmentedconditionallikelihood2} L_{2, \mathrm{com}}( \Sigma) =
\prod_{i=0}^n
\frac{ \exp (-({1}/{2})\sum_{j \in\mathcal{J}_i}
\mathbf{1}[M_i = j] W_{ij}' \Sigma^{-1} W_{ij}
) } {
\sigma\sum_{l \in\mathcal{J}_i} \exp( -V_{il}' \Sigma_{11}^{-1}
V_{il} / 2) }.
\end{equation}

The term $\sum_{l \in\mathcal{J}_i} \exp(-V_{il}'\Sigma
_{11}^{-1}V_{il}/2)$ can be viewed as an approximation of a Gaussian
integral; if we assume that, for all $i, \{V_{il}, l \in\mathcal{J}_i\}
$ resemble $|\mathcal{J}_i|$ i.i.d. draws from some distribution $V
\sim\N(0, \Omega)$, then we have
%
%e3.10 #&#
\begin{eqnarray}\label{Vapproximation}
&& \sum_{l \in\mathcal{J}_i} \exp\bigl(
-V_{il}' \Sigma_{11}^{-1}
V_{il} / 2 \bigr)\nonumber
\\
&&\qquad \approx |\mathcal{J}_i| \int\exp
\bigl( -V' \Sigma_{11}^{-1} V/2 \bigr)
\frac{ \exp( -V' \Omega^{-1} V/2 ) } {
(2\pi)^{p/2} | \Omega|^{1/2} } \,dV
\nonumber\\[-8pt]\\[-8pt]
&&\qquad =  |\mathcal{J}_i| \biggl( \frac{ | ( \Sigma_{11}^{-1} + \Omega^{-1}
)^{-1} | } {|\Omega| }\biggr)^{1/2}\nonumber
\\
&&\qquad = |\mathcal{J}_i| \biggl( \frac{ |\Sigma_{11}| }{ |\Sigma_{11} +
\Omega| }\biggr)^{1/2}.\nonumber
\end{eqnarray}
Noting that $\sigma|\Sigma_{11}|^{1/2} = |\Sigma|^{1/2}$, and ignoring
multiplicative constants, we arrive at an approximate augmented-data
conditional likelihood:
\[
L_{2, \mbox{com}}( \Sigma) \approx\biggl(
\frac{ | \Sigma_{11} + \Omega| }{ |\Sigma| } \biggr)^{(n+1)/2} \exp
\Biggl( -\frac{1}{2} \sum
_{i=0}^n \sum_{j \in\mathcal{J}_i}
\mathbf{1}[M_i = j] W_{ij}'
\Sigma^{-1} W_{ij} \Biggr).
\]
Typically $\Sigma_{11} \ll\Omega$, meaning
\begin{eqnarray*}
\partial\bigl( \log\bigl(|\Sigma_{11} + \Omega|\bigr) - \log\bigl(|\Sigma|\bigr) \bigr) &
= & \operatorname{Tr}\bigl((\Sigma_{11} + \Omega)^{-1}\,
\partial\Sigma_{11}\bigr) - \operatorname{Tr}\bigl(
\Sigma^{-1} \,\partial\Sigma\bigr)
\\
& \approx& - \operatorname{Tr}\bigl( \Sigma^{-1} \,\partial\Sigma
\bigr)
\end{eqnarray*}
as $\partial\log(|\Sigma|)$ dominates $\partial\log(|\Sigma_{11} +
\Omega|)$. This justifies the approximation $\log(|\Sigma_{11} + \Omega
|) - \log(|\Sigma|) \approx-\log(|\Sigma|)$ in the augmented-data
conditional log-likelihood, as it will admit nearly the same maximizer.
Thus, we have
%
%e3.11 #&#
\begin{eqnarray}\label{approximateACL2}
\log\bigl(L_{2,\mathrm{com}}(\Sigma)\bigr)
&\approx& -
\frac{n+1}{2} \log\bigl( | \Sigma| \bigr)
\nonumber\\[-8pt]\\[-8pt]
&&{}  - \frac{1}{2} \sum
_{i=0}^n \sum_{j \in\mathcal{J}_i}
\mathbf{1}[M_i = j] W_{ij}'
\Sigma^{-1} W_{ij}.\nonumber
\end{eqnarray}

While \eqref{approximateACL2} is much easier to work with than the
logarithm of the exact conditional likelihood \eqref
{augmentedconditionallikelihood2}, the assumptions of this
approximation are somewhat \mbox{tenuous}. Under this model \eqref{DGP}, both
conditional and marginal distributions of observations at each time
point follow a normal mixture, meaning for $l$ randomly chosen from
$\mathcal{J}_i$, we have a difference of normal mixtures (itself a
normal mixture) for $V_{il}$, instead of i.i.d. normals as \eqref
{Vapproximation} suggests. We nevertheless proceed with approximation
\eqref{approximateACL2} in place of \eqref
{augmentedconditionallikelihood2}, noting that convergence of the EM
algorithm needs to be more carefully monitored in this instance.

At each iteration of the EM algorithm, we maximize the so-called $Q$ function:
%
%e3.12 #&#
\begin{eqnarray}
\label{Qfun} Q^{(t)}( \Sigma) &=& \mathbb{E}_{\Sigma^{(t)}}\bigl[
\log\bigl( L_{2,\mathrm{com}}( \Sigma) \bigr) | Y \bigr]
\nonumber\\[-8pt]\\[-8pt]
&\approx& - \frac{n+1}{2} \log\bigl( |\Sigma| \bigr) - \frac{1}{2} \sum
_{i=0}^n \sum_{j \in\mathcal{J}_i}
\omega_{ij} W_{ij}' \Sigma^{-1}
W_{ij},\nonumber
\end{eqnarray}
with $\Sigma^{(t)} = \operatorname{argmax}(Q^{(t-1)}(\Sigma))$ and
$\omega_{ij} = \mathbb{E}_{\Sigma^{(t)}} [ \mathbf{1}[ M_{i} = j ] | Y
]$. Clearly,
\[
\omega_{ij} = \frac{ \exp( -W_{ij}' [ \Sigma^{(t)} ]^{-1} W_{ij} / 2 )
} {
\sum_{l \in\mathcal{J}_i} \exp( -W_{lj}' [ \Sigma^{(t)} ]^{-1} W_{lj}
/ 2 ) }.
\]
The maximizer of \eqref{Qfun} can be found in closed form as a weighted
sample covariance matrix,
%
%e3.13 #&#
\begin{equation}
\label{Qsol} \Sigma^{(t+1)} = \frac{1}{n+1} \sum
_{i=0}^n \sum_{j \in\mathcal{J}_i}
\omega_{ij} W_{ij} W_{ij}'.
\end{equation}

Again, due to several different approximations used in maximizing the
original conditional likelihood \eqref{conditionallikelihood}, it is
necessary to monitor the convergence to a suitable (if slightly
suboptimal) solution, as the log-likelihood is not guaranteed to
increase at each iteration.

%s3.4 #&#
\subsection{A prediction model for fast implementation}\label{subsecprediction}

Exact closed-form expressions for \mbox{$k$-}step ahead predictive
distributions are not available for the model \eqref{DGP}. Because of
the need for real-time forecasting of many steps ahead, we explore
approximations to \mbox{$k$-}step ahead predictive distributions that are
available in closed form. An immediate approach to doing so is to
explore whether the approximate complete-data conditional
log-likelihood used for inference \eqref{approximateACL2} corresponds
to a probabilistic model (perhaps misspecified) that admits closed-form
predictive distributions. In other words, if the previous section
derives an approximate log-likelihood \eqref{approximateACL2} from an
exact model \eqref{DGP}, here we treat \eqref{approximateACL2} as exact
and explore corresponding approximate models.

Let $Z_i = (Y_{i-p} \cdots Y_{i-1} Y_i)'$ for $0 \leq i \leq n$. Since
$W_{ij} = Z_i - Z_j$, we may arrive at the likelihood expression \eqref
{approximateACL2} by assuming $Z_i \sim\N(Z_{i-M_i}, \Sigma)$
independently. This is obviously a misspecification, since for any $k
\leq p$, $Z_i$ and $Z_{i+k}$ contain duplicate entries and thus cannot
be independent. But assuming the $\{Z_i\}$ independent, and further
assuming $\mathbb{P}(M_i = j) = 1/|\mathcal{J}_i|$ independently for
all $i$, we can\vadjust{\goodbreak} write the (conditional) likelihood for an independent
multivariate normal mixture model, denoted $L_a$ to distinguish from $L_{2,
\mathrm{com}}$:
%
%e3.14 #&#
\begin{equation}
\label{mixturelikelihood} L_a(\Sigma) = \prod_{i=0}^n
\prod_{j \in\mathcal{J}_i} \biggl[ | \Sigma|^{-1/2} \exp
\biggl( -\frac{1}{2} W_{ij}' \Sigma^{-1}
W_{ij} \biggr) \biggr]^{\mathbf{1}[M_i = j]}.
\end{equation}

Indeed, we see that $L_a(\Sigma)$ is equal to the approximation of
$L_{2,\mathrm{com}}(\Sigma)$ given in~\eqref{approximateACL2}. Thus,
the misspecified independent mixture model for $Z_i$ yields the same
likelihood ($L_a$) as the approximation to $L_2$, the exact
(conditional) likelihood corresponding to the data-generating process.
Also, recall that $M_i = j$ denotes $Z_i$ as an instance of motif
$Z_j$. The implied relation in \eqref{mixturelikelihood} is that
%
%e3.15 #&#
\begin{equation}
\label{relation2} Z_j \rightarrow Z_i\qquad\mbox{if }
Z_i | Z_j \sim\N(Z_j, \Sigma)
\end{equation}
and, indeed, this relation is closely connected to the one defined in
\eqref{fullconditional}. They appear equivalent, as \eqref
{fullconditional} is recovered by assuming $Z_i|Z_j \sim\N(Z_j, \Sigma
)$, and then considering the conditional distribution $Y_i|Y_{(-m)\dvtx
(i-1)}$. However, for \eqref{relation2} to hold for all $i$ requires
the impossible assumption of $Z_i$ being independent of $Z_{i-1}$,
while the relation in \eqref{fullconditional} does not.

The corresponding $Q$ function for this complete-data conditional
likelihood \eqref{mixturelikelihood} is
\[
Q_a^{(t)}(\Sigma) = \sum
_{i=0}^n -\frac{1}{2} \log\bigl( | \Sigma| \bigr) -
\frac{1}{2} \sum_{j \in\mathcal{J}_i} \mathbb{E}_{\Sigma^{(t)}}
\bigl[\mathbf{1}[M_i = j] | Z\bigr] W_{ij}'
\Sigma^{-1} W_{ij}.
\]
Working $\mathbb{E}_{\Sigma^{(t)}}[\mathbf{1}[M_i = j] | Z] = \omega
_{ij}$, we see that $Q_a^{(t)}$ is identical to $Q^{(t)}$ given in
\eqref{Qfun}, confirming that the ``same'' $\Sigma$ parametrizes both
the original data-generating process assumed in \eqref{DGP} and its
degenerate approximation that we will use to make predictions in \eqref
{mixturelikelihood}. We may also think of maximizing $Q$ as inferring
motif instances given by the relation \eqref{relation2}, that is,
minimizing a distance metric.

The independent multivariate mixture distribution of $\{Z_i\}$
considered here very easily provides \mbox{$k$-}step predictive distributions
for $k \leq p$. If we have observed the process up to $Y_n$ and wish to
predict $Y_{n+k}$, this is equivalent to having observed $Z$ up to
$Z_n$ and wishing to predict the last component of $Z_{n+k}$. Having
observed $Z_n$ completely, we have observed the first $p-k+1$
components of $Z_{n+k}$, and thus by the (misspecfied) independence
assumed for $\{Z_i\}$, the predictive distribution for $Y_{n+k}$
depends only on these $p-k+1$ values. To write this, we denote $\tilde
{Z}^{k}_n$ as the first $p-k+1$ components\vspace*{1pt} of $Z_{n+k}$ (or the last
$p-k+1$ components of $Z_n$); also, let $\tilde{W}^k_{nj} = \tilde
{Z}^k_n - \tilde{Z}^k_j$ and partition $\Sigma$ into $\Sigma^k_{11}$ as
the upper-left $(p-k+1) \times(p-k+1)$ submatrix, $\Sigma^k_{22}$ as
the single bottom-right element (thus identical to $\Sigma_{22}$), and
$\Sigma^k_{12}, \Sigma^k_{21}$ accordingly. Then we have
%
%e3.16 #&#
\begin{equation}
\label{predictivedistn} Y_{n+k}|Y_{(-m)\dvtx n} \sim\sum
_{j \in\mathcal{J}_{n+k}} \alpha^{k}_j \N\bigl(
\mu^{k}_j, \sigma^2_k \bigr),\vadjust{\goodbreak}
\end{equation}
where:
\begin{itemize}
\item$\alpha^{k}_j = \mathbb{P}(M_{n+k} = j | \tilde{Z}^{k}_n) \propto
\exp(- (\tilde{W}^k_{nj})' [\Sigma^k_{11}]^{-1} \tilde{W}^k_{nj} / 2 )$,\vspace*{2pt}
\item$\mu^k_j = Y_{n+k-j} + \Sigma^k_{21}[\Sigma^k_{11}]^{-1}\tilde{W}^k_{nj}$,\vspace*{2pt}
\item$\sigma^2_k = \Sigma^k_{22} - \Sigma^k_{21} [\Sigma^k_{11}]^{-1}
\Sigma^k_{12}$.
\end{itemize}

In terms of motifs, these predictive distributions result from
considering the most recent subseries of the data of length $p-k+1$ as
a partially observed motif instance, $Z_{n+k}$, which includes the
future observation we wish to predict, $Y_{n+k}$. Using the implied
motif relation in \eqref{relation2}, we infer both the motif for which
$Z_{n+k}$ is an instance and derive predictive distributions using
simple multivariate normal properties \eqref{predictivedistn}.

Of course, we use $\hat{\Sigma}$, the solution to \eqref{Qsol}, in
place of $\Sigma$ in the above expressions, acknowledging that the
resulting predictive distributions fail to account for the uncertainty
in our estimate of $\Sigma$.

%s3.5 #&#
\subsection{Interpreting \texorpdfstring{$\hat{\Sigma}$}{hat Sigma}} \label{subsecinterpreting}

\begin{figure}
\begin{tabular}{@{}c@{\qquad}c@{}}

\includegraphics{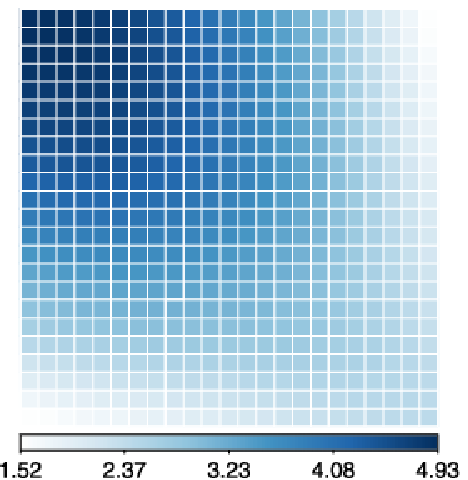}
 & \includegraphics{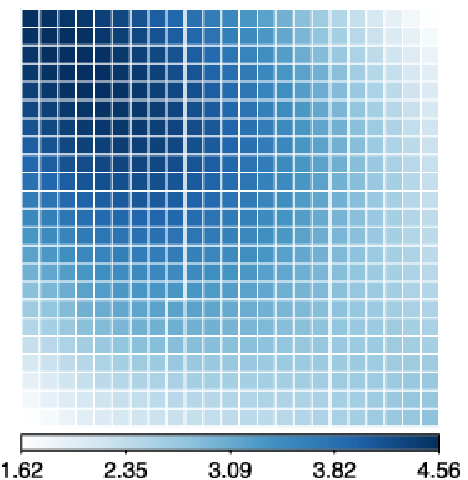}\\
\footnotesize{(A) $\hat{\Sigma}$ for patient 10, day 1, beam 1}%\label{figsub1}
&  \footnotesize{(B) $\hat{\Sigma}$ for patient 9, day 1, beam 2} %\label{figsub2}
\end{tabular}
\caption{Illustration of $\hat{\Sigma}$ for two of the time series in
our data, using $p=22$. Note that the color scale differs slightly for
each figure.}\label{sigma}
\end{figure}

Figure~\ref{sigma} shows estimates $\hat{\Sigma}$ from two of the time
series in our data. Interpreting these as covariance matrices, we see
relatively high correlations across components, favoring instantiating
motifs where the difference between the motif instance and the original
motif is roughly linear with a slope near 0. Also, the diagonal terms
are decreasing from top to bottom, implying that more weight is given
to the most recent components of the observed time series when
inferring the latent motif instance and making predictions.

%
%f6 #&#

%s4 #&#
\section{Evaluating out-of-sample prediction error with competing methods}\label{secevaluation}

We compare out-of-sample prediction performance for tumor tracking
using the LMAR($\Sigma$) model with three methods that are
straightforward to implement and provide real-time forecasts. Neural
networks (\ref{subsecNN}) and ridge regression (\ref{subsecridge}) both
compare favorably to alternative methods with regards to prediction
accuracy [\citet{sharpprediction2004,krausscomparative2011}].
LICORS (\ref{subseclic}) is a nonparametric and nonregression
forecasting method based on predictive state representations of the
time series [\citeauthor{licors} (\citeyear{licors,mixedlicors})]. For each method,
Sections~\ref{subsecpreprocessing}--\ref{subseccomputation} discuss
data preprocessing and computational considerations relevant for
real-time tumor tracking.

%s4.1 #&#
\subsection{Feedforward neural networks} \label{subsecNN}
Multilayer feedforward neural networks with at least one hidden layer
have been used to forecast lung tumor motion by \citet
{murphy2002adaptive} and \citet{murphycomparative2006}, as well as
in simultaneous comparisons of several methods [\citet
{sharpprediction2004,krausscomparative2011,ernst2013evaluating}]. Using
$p \times h \times1$ neural networks, we can predict $Y_{i+k}$ as a
function of $Y_{(-m)\dvtx i}$. Let $X_i = Y_{(i-p) + 1\dvtx p}$, then
%
%e4.1 #&#
\begin{equation}
\label{NN} \hat{Y}_{i+k} = \beta_0 +
\beta'G(X_i),
\end{equation}
where $G(X_i)=(g(w_{01} + w_1'X_i) g(w_{02} + w_2'X_i) \cdots g(w_{0h}
+ w_h'X_i))'$ with activation function $g$; here we assume $g(x) =
1/(1+\exp(-x))$. Hyperparameters $p$ and $h$ are set by the user (as is
the form of the activation function). Unknown parameters $\beta_0, \beta, w_{01}, \ldots, w_{0h}, w_1, \ldots, w_h$ are estimated by minimizing
the sum of squares using the \texttt{R} package \texttt{nnet}
[\citet{ripley}]. Because the number of unknown parameters is
large ($w_1, \ldots, w_h$ are $p$-vectors), to prevent overfitting, a
regularization term is often used in the sum of squares minimization.
Then, the model is fit by minimizing
%
%e4.2 #&#
\begin{equation}
\label{cost} C(Y,\theta) = \sum_{i=0}^{n-k}
(\hat{Y}_{i+k} - Y_{i+k})^2 + \lambda
\theta'\theta,
\end{equation}
where $\theta$ represents a vector of all unknown parameters stacked
together and $\lambda$ is a penalty hyperparameter that is supplied by
the user, with higher values providing more shrinkage.

%s4.2 #&#
\subsection{Ridge regression}
\label{subsecridge}
The second competing method considered is a linear predictor of the form
%
%e4.3 #&#
\begin{equation}
\label{ridge} \hat{Y}_{i+k} = \beta_0 +
\beta'X_i,
\end{equation}
with $X_i = Y_{(i-p) + 1\dvtx p}$ and where $\beta_0, \beta$ are found
by minimizing
%
%e4.4 #&#
\begin{equation}
\label{cost2} C(Y, \beta_0, \beta) = \sum
_{i=0}^{n-k} (\hat{Y}_{i+k} -
Y_{i+k})^2 + \lambda\bigl(\beta_0^2
+ \beta'\beta\bigr).
\end{equation}
Nearly all studies involving forecasting lung tumor motion consider
predictors of this form, usually referred to as ridge regression.
However, since ridge regression assumes $\{Y_i\}$ to be independent
[\citet{hoerl1970ridge}], the model implied by \eqref
{ridge}--\eqref{cost2} is better described as fitting an autoregressive
model of order $p+k-1$ (the first $k-1$ coefficients being 0) using
conditional least squares, with an $L_2$ penalty on the vector of
autoregressive coefficients (yet we shall refer to this prediction
method as ridge regression). Linear models lack many features that seem
appropriate for this forecasting example, such as multimodal and/or
heteroskedastic conditional distributions, yet still perform reasonably
well and are commonly used as a baseline for comparing tumor prediction methods.

%s4.3 #&#
\subsection{Light cone reconstruction of states (LICORS)} \label{subseclic}
Mixed LICORS [\citet{mixedlicors}] is a recent nonparametric
forecasting method based on predictive state representations of
spatiotemporal fields [\citet{shalizi2003optimal,licors}]. In the
context of our forecasting example, mixed LICORS models
$Y_{i+k}|Y_{(-m)\dvtx i}$ as depending only on the \textit{past light
cone} (with horizon $p$) $X_i = Y_{(i-p) + 1\dvtx p}$; furthermore,
$\varepsilon(X_i)$ is a minimal sufficient statistic for the predictive
distribution of $Y_{i+k}$, so that
%
%e4.5 #&#
\begin{equation}
\label{licors} Y_{i+k}|Y_{(-m)\dvtx i} \sim Y_{i+k}|X_i
\sim Y_{i+k}|\varepsilon(X_i),
\end{equation}
and if $\varepsilon(X_i) = \varepsilon(X_j)$, then $Y_{i+k}|\varepsilon
(X_i) \sim Y_{j+k}|\varepsilon(X_j)$. Without loss of generality, we
may assume $\varepsilon$ takes values in $\mathcal{S} = \{s_1, \ldots,
s_K\}$, and for simpler notation let $S_i = \varepsilon(X_i)$ and
denote $\P_j(Y_{i+k}) = \P(Y_{i+k}|S_i = s_j)$. The unknown parameters
of this model are the mapping $\varepsilon$, the number of predictive
states $K$ and the predictive distributions of the predictive states $\{
\P_j, 1 \leq j \leq K\}$. For fixed $K$, the remaining parameters are
estimated by maximizing
%
%e4.6 #&#
\begin{equation}
\label{cost3} C(Y, \varepsilon, \P_1, \ldots, \P_K) =
\prod_{i=0}^{n-k} \sum
_{j=1}^K \P_j(Y_{i+k})
\P(S_i = j|X_i),
\end{equation}
which acts as a likelihood, except for $\P_j$ being unknown. \citet
{mixedlicors} maximized \eqref{cost3} with a nonparametric variant of
the EM algorithm using weighted kernel density estimators to
approximate the unknown densities of the predictive distributions $\{\P
_j, 1 \leq j \leq K\}$; they also advocated data-driven procedures for
choosing the number of predictive states $K$.

It is possible to embed the LMAR model in a parametric (Gaussian) mixed
LICORS framework, treating $\{V_{ij}, j \in\mathcal{J}_i\}$ as the
past light cone $\ell_i^-$ and $\{V_{ij}$
where $M_i = j\}$ as the predictive state $S_i = \varepsilon(\ell_i)$.
While this choice of $\varepsilon$ does provide a minimal sufficient
statistic for the predictive distribution of $Y_i$ (or $L_i^+$) under
the LMAR model, it will not provide any dimension reduction or
parsimony since $\varepsilon(\ell_i)$ will almost surely be unique for
each $i$ under our model assumptions.

Fitting the mixed LICORS model to the time series in our data and using
it for forecasting was accomplished using the \texttt{R} package \texttt
{LICORS} [\citet{Rlicors}]. Note that point forecasts using the
inferred model \eqref{licors} will be a weighted average of the means
of the predictive states $s_i \in\mathcal{S}$.

%s4.4 #&#
\subsection{Data preprocessing}\label{subsecpreprocessing}

Similar to \citet{krausscomparative2011}, we use a total of 80
seconds of data (2400 observations) from each time series, 40 seconds
for model fitting and 40 seconds for out-of-sample prediction given the
model fit to the first 40 seconds of data. This necessitates removing
time series for which we have fewer than $2400+k$ observations, where
$k$ is the forecast window. This eliminates 61 of the 171 time series
in our data base, unfortunately including all time series from patients
1, 2 and 3. An additional 15 time series were eliminated because there
were several gaps in the observation sequence. This leaves us with 95
total time series; patient 8 has only one time series and patient 6 has
the next fewest series with 9. Patient 11 has the most time series with
21. While each time series is three dimensional, we predict using only
the first principal component (the principal component transformation
is estimated from the initial 40~s of training data) as discussed in
Section~\ref{subsecfeatures}.
%DC\dvtx We could just have looked later on in the time series, picking
%some time after a gap as t=0.

%s4.5 #&#
\subsection{Tuning hyperparameters}\label{subsectuning}

Because of the need for real-time model fitting and prediction, all
tuning and hyperparameters for the methods we consider must be
specified prior to the administration of radiotherapy---before any data
is observed. This suggests finding specifications for each model that
perform reasonably well for all patients, though perhaps sub-optimally
for each patient individually. Indeed, this is the approach usually
taken in the literature [\citet
{sharpprediction2004,krausscomparative2011,ernst2013evaluating}].
Because patients are typically given several or many instances of
radiotherapy during different sessions, there seems to be potential for
more patient-specific tuning of hyperparameters, though this is left as
a separate problem for now.

%
%t2 #&#
\begin{table}
\tabcolsep=0pt
\caption{List of global, patient-independent hyperparameters to be tuned for each prediction method}\label{hyperparamtable}
\begin{tabular*}{\tablewidth}{@{\extracolsep{\fill}}@{}lcc@{}}
\hline
\textbf{Method} & \textbf{Hyperparameter} & \textbf{Description}\\
\hline
LMAR & $p$ & Motif length \eqref{relation2}
\\[3pt]
Neural networks & $p$ & Length of input vector $X_i$ \eqref{NN} \\
& $h$ & Number of neurons in hidden layer \eqref{NN} \\
& $\lambda$ & Shrinkage; L2 penalty \eqref{cost}
\\[3pt]
Ridge regression & $p$ & Length of input vector $X_i$ \eqref{ridge} \\
& $\lambda$ & Shrinkage; L2 penalty \eqref{cost2}
\\[3pt]
Mixed LICORS & $p$ & Length of input vector $X_i$ \eqref{licors} \\
\hline
\end{tabular*}
\end{table}

Table~\ref{hyperparamtable} lists the hyperparameters and/or tuning
parameters for each of the prediction methods we consider. As described
in Section~\ref{subsecpreprocessing}, since the first 40 seconds of
each time series will not be used to evaluate out-of-sample prediction,
we may use these subseries to find sensible, patient-independent values
for all hyperparameters. Each 40 second subseries is further divided,
where for a given set of hyperparameters each prediction method is fit
to the first 30 seconds of data (900 observations), and then the
remaining 10 seconds are used to generate out-of-sample predictions,
for which we store the vector of errors.

Using a course grid search over the parameter space given in Table~\ref{hyperparamtable}, predictive error [both root mean squared error
(RMSE) as well as median absolute error (MAE), which is more robust to
heavy-tailed error distributions] is averaged across patients, allowing
us to choose the best set of patient-independent hyperparameter values
[\citet{krausscomparative2011}]. Note that different
hyperparameter values are chosen for different forecast windows.

%s4.6 #&#
\subsection{Computational considerations}\label{subseccomputation}

In addition to providing real-time forecasts, tumor tracking models
require parameters that can be estimated very quickly so that accurate
(forecast-assisted) radiotherapy can begin as soon as possible after
observing a short window of training data.

Ridge regression yields almost instantaneous estimates of parameters
necessary for prediction [$\beta$ in \eqref{ridge}], since \eqref
{cost2} can be minimized in closed form. Fitting neural networks \eqref
{NN}, however, requires numerical optimization of \eqref{cost}. This
was carried out using the \texttt{nnet} package in \texttt{R}, which
implements the BFGS algorithm [\citet{ripley}]. Because \eqref
{cost} is not convex, we recommend several random starting points for
initiating the optimization, insomuch as time allows; the dimension of
the parameter space and the convergence criteria for the numerical
optimization are both extremely important considerations in addition to
the length of the time series being fit. For example, on a Lenovo X220
laptop with an Intel Core i5-2520~M 2.50 GHz processor, a $45 \times6
\times1$ neural network required about 10 seconds to fit on 1200
observations when using \texttt{nnet}'s default convergence criteria,
with 10 randomly initialized starting points.

The computation time in fitting the LMAR($\Sigma$) depends critically
on both the convergence criteria for the EM algorithm as well as the
initial value of $\Sigma$ used. Typically, the likelihood \eqref
{conditionallikelihood} or log-likelihood is used, however, the EM
\mbox{updates} given in \eqref{Qsol} are only approximate, meaning the
likelihood is not guaranteed to increase at every iteration. We found
that using the approximate log-likelihood~\eqref{approximateACL2} to
check convergence yielded convergence in the exact \mbox{log-}likelihood. This
being the case, other metrics could possibly be used to check
convergence that are quicker to calculate than \eqref{approximateACL2},
such as the Frobenius norm of differences in the updates of $\hat{\Sigma
}$. To obtain good starting values, the algorithm can be run before
having observed the entire training sequence using a simple starting
value of a diagonal matrix. Using a relative tolerance of 0.0001 for
the approximate log-likelihood, we were able to compute $\hat{\Sigma}$
in no more than four seconds for each of the time series considered.
\texttt{R} code for fitting the LMAR model is included in this paper's
supplementary materials [\citet{cervone-code}].

The value of $m$ for the LMAR model may also trade off estimation speed
and accuracy; we used $m=400$, though found essentially identical
results for $m=200$ and $m=300$ (higher values of $m$ favor faster, but
less precise, estimation of $\Sigma$).

Parameter estimation for mixed LICORS took several minutes on our
machine. However, much of this computational cost is accrued in
inferring $K$, the number of predictive states. The procedure described
in \citet{mixedlicors} and implemented in the \texttt{LICORS}
\texttt{R} package is to start at an upper bound for the number of
predictive states, optimize the likelihood approximation \eqref{cost3}
and then merge the two states whose predictive distributions are
closest (measured by some distance or a hypothesis test). The
optimizing and merging steps are repeated until we either have 1 state
remaining or, alternatively, all pairwise tests for equality among
predictive distributions are rejected. Then, cross-validation is used
to choose among these candidate models indexed by different values of $K$.

While there may be some loss in prediction accuracy, estimation speed
can be improved by fixing $K$ (perhaps tuning it as in Section~\ref{subsectuning}). Furthermore, initializing the nonparametric EM
algorithm with informative starting values (learned from previously
observed respiratory trace curves) and relaxing the convergence
criteria may substantially increase estimation speed with little loss
in predictive performance.

%s5 #&#
\section{Prediction results for tumor tracking data}\label{secresults}
The results of out-of-sample predictions using the LMAR model, as well
as the methods discussed in Section~\ref{secevaluation}, are provided
in this section. Point forecasts are discussed in Sections~\ref{subsecpointresults}--\ref{subsecqualsummaries} and
interval/distributional forecasts in Section~\ref{subsecintervals}.

%s5.1 #&#
\subsection{Results for point forecasts}\label{subsecpointresults}

The measures of predictive performance we consider are root mean
squared error (RMSE) and median absolute error (MAE), as well as the
fraction of time each forecasting method obtains the minimum prediction
error among the methods compared. We report these quantities for each
of the 8 patients, at forecast windows of 0.2~s (6 observations), 0.4~s
(12 observations) and 0.8~s (18 observations) in Table~\ref{resultspoint}.

%
%t3 #&#
\begin{table}[t]
\caption{Summary of errors in point forecasts for all four methods and
all three forecast windows considered. RMSE is root mean squared error,
MAE is median absolute error, and \emph{Best} refers to the proportion
of time for which the absolute prediction error is smallest among the
methods considered. For each metric, the most desirable value among the
four methods for each patient/forecast window combination is in \textbf{bold}}\label{resultspoint}
\tabcolsep=0pt
\begin{tabular*}{\tablewidth}{@{\extracolsep{\fill}}@{}lcccccccccc@{}}
\hline
& & \multicolumn{3}{c}{\textbf{0.2~s forecast}} & \multicolumn{3}{c}{\textbf{0.4~s forecast}} & \multicolumn{3}{c@{}}{\textbf{0.6~s forecast}}\\[-6pt]
& & \multicolumn{3}{c}{\hrulefill} & \multicolumn{3}{c}{\hrulefill} & \multicolumn{3}{c@{}}{\hrulefill}\\
\textbf{Patient} & \textbf{Method} & \textbf{RMSE} & \textbf{MAE} & \textbf{Best} & \textbf{RMSE} & \textbf{MAE} & \textbf{Best} & \textbf{RMSE} & \textbf{MAE} & \textbf{Best}\\
\hline
\phantom{0}4 & LMAR & 0.52 & 0.24 & 0.27 & 0.99 & 0.39 & 0.27 & \textbf{1.18} & \textbf{0.44} & \textbf{0.31} \\
& NNs & \textbf{0.46} & \textbf{0.22} & \textbf{0.28} & \textbf{0.90} & 0.39 & \textbf{0.28} & 1.20 & 0.48 & 0.27 \\
& Ridge & 0.53 & 0.31 & 0.20 & 1.08 & 0.62 & 0.17 & 1.50 & 0.86 & 0.18\\
& LICORS & 0.58 & 0.25 & 0.25 & 1.05 & \textbf{0.37} & \textbf{0.28} & 1.43 &0.52 & 0.24
\\[3pt]
\phantom{0}5 & LMAR & 0.56 & \textbf{0.25} & \textbf{0.30} & 0.96 & 0.42& 0.29 & \textbf{1.15} & \textbf{0.51} & \textbf{0.30} \\
& NNs & \textbf{0.55} & 0.27 & 0.27 & \textbf{0.89} & \textbf{0.40} & \textbf{0.30} &\textbf{1.15} & \textbf{0.51} & \textbf{0.30} \\
& Ridge & 0.58 & 0.31 & 0.25 & 1.01 & 0.56 & 0.23 & 1.39 & 0.78 & 0.23\\
& LICORS & 0.79 & 0.35 & 0.19 & 1.33 & 0.63 & 0.18 & 1.79 & 0.89 &0.17
\\[3pt]
\phantom{0}6 & LMAR & \textbf{0.77} & \textbf{0.40} & \textbf{0.29} & \textbf{1.54} & \textbf{0.82} & \textbf{0.30} & \textbf{2.00} & \textbf{1.06} & \textbf{0.34} \\
& NNs & 1.01 & 0.46 & 0.24 & 1.74 & 0.93 & 0.24 & 2.43 & 1.38 & 0.22 \\
& Ridge & 0.83 & 0.42 & 0.28 & 1.59 & 0.88 & 0.28 & 2.14 & 1.28 & 0.28\\
& LICORS & 1.37 & 0.57 & 0.19 & 2.17 & 1.19 & 0.18 & 2.92 & 1.75 &0.15
\\[3pt]
\phantom{0}7 & LMAR & \textbf{0.40} & \textbf{0.15} & \textbf{0.35} & \textbf{0.85} & \textbf{0.27} & \textbf{0.37} & \textbf{1.23} & \textbf{0.41} & \textbf{0.36} \\
& NNs & 0.43 & 0.19 & 0.26 & 0.88 & 0.36 & 0.25 & 1.35 & 0.51 & 0.25 \\
& Ridge & 0.44 & 0.26 & 0.20 & 1.00 & 0.59 & 0.16 & 1.56 & 0.96 & 0.17\\
& LICORS & 0.62 & 0.25 & 0.20 & 1.05 & 0.41 & 0.21 & 1.56 & 0.56 &0.23
\\[3pt]
\phantom{0}8 & LMAR & 1.27 & \textbf{0.62} & \textbf{0.27} & \textbf{2.63} &1.46 & 0.26 & 3.57 & 2.00 & 0.24 \\
& NNs & \textbf{1.26} & 0.68 & \textbf{0.27} & 2.71 & \textbf{1.27} & \textbf{0.28} &\textbf{3.46} & \textbf{1.76} &\textbf{ 0.29} \\
& Ridge & 1.44 & 0.69 & 0.20 & 2.86 & 1.54 & 0.19 & 4.11 & 2.26 & 0.19\\
& LICORS & 1.50 & 0.64 & 0.26 & 2.89 & 1.33 & \textbf{0.28} & 3.70 & \textbf{1.76} & 0.28
\\[3pt]
\phantom{0}9 & LMAR & \textbf{0.58} & \textbf{0.22} & \textbf{0.39} & \textbf{1.29} & \textbf{0.52} & \textbf{0.35} & \textbf{2.03} & \textbf{0.90} & \textbf{0.30} \\
& NNs & 0.73 & 0.32 & 0.24 & 1.69 & 0.64 & 0.26 & 2.45 & 0.92 & 0.24 \\
& Ridge & 0.81 & 0.34 & 0.22 & 1.68 & 0.73 & 0.22 & 2.42 & 0.98 & 0.25\\
& LICORS & 1.35 & 0.53 & 0.15 & 2.20 & 0.98 & 0.17 & 2.64 & 1.19 &0.20
\\[3pt]
10 & LMAR & \textbf{0.88} & \textbf{0.36} & \textbf{0.34} & \textbf{1.73} & \textbf{0.77} & \textbf{0.33} & \textbf{2.55} & \textbf{1.19} & \textbf{0.30} \\
& NNs & 1.09 & 0.44 & 0.25 & 2.16 & 0.93 & 0.24 & 2.98 & 1.35 & 0.24 \\
& Ridge & 0.95 & 0.45 & 0.24 & 1.84 & 0.94 & 0.24 & 2.67 & 1.41 & 0.26\\
& LICORS & 1.62 & 0.61 & 0.17 & 2.20 & 1.10 & 0.19 & 3.25 & 1.56 &0.20
\\[3pt]
11 & LMAR & \textbf{1.13} & \textbf{0.44} & \textbf{0.32} & \textbf{2.59} & \textbf{1.06} & \textbf{0.29} & \textbf{3.70} & \textbf{1.49} & \textbf{0.31} \\
& NNs & 1.24 & 0.50 & 0.25 & 2.95 & 1.19 & 0.24 & 3.99 & 1.70 & 0.23 \\
& Ridge & 1.19 & 0.63 & 0.22 & 2.69 & 1.51 & 0.21 & 3.99 & 2.40 & 0.21\\
& LICORS & 1.64 & 0.57 & 0.21 & 3.04 & 1.09 & 0.26 & 4.21 & 1.65 &0.25 \\
\hline
\end{tabular*}
\end{table}

We stress that RMSE may not be the most useful summary of predictive
performance since the error distributions are heavy tailed, and in the
application of radiotherapy, we are more concerned with whether or not
the treatment beam was localized to the tumor than with the squared
distance of the treatment beam to the tumor.\footnote{However, the loss
function implied in the model fitting and point prediction is squared
error loss, which is the simplest for many computation reasons.} For
this reason, we feel that the median (more generally, quantiles of the
distribution function for absolute errors) is the best summary of
predictive performance for this data context. Ultimately, the
dosimetric effects of these errors are of most interest, but their
determination is complicated and beyond the scope of this work.

Two further points of emphasis regarding the accuracy summaries are
that while we eliminated time series with unevenly spaced observations
from consideration, we still have quite a few time series with unusual
motion in our data base. Without actually observing the patient, we are
not sure whether observed deviations from normal breathing are caused
by exogenous factors or are instances of relevant components of the
data-generating process, such as coughs, yawns, deep breaths, etc. The
other point is that there is a lot of disparity in the measures of
predictive performance within the literature on this subject; in
addition to working with different data sets, obtained from differing
equipment, some authors account for the between-patient variation in
respiratory dynamics by scaling or normalizing all curves or by
comparing errors from a prediction method against errors from making no
prediction and just using the lagged value of the series. When using
evaluation procedures of \citet{krausscomparative2011} and
\citet{murphycomparative2006}, we produced very similar results
with ridge regression and linear models. However, the error summaries
we present here, in comparison with the LMAR model, are not directly
comparable to these results.

%s5.2 #&#
\subsection{Quantitative summaries of point forecasts}\label{subsecquantsummaries}
Summarizing Table~\ref{resultspoint}, we see that ridge regression is
actually suboptimal in all accuracy measures for all patients and
forecast windows. The LMAR model strongly outperforms the other three
methods for all forecast windows for patients 6, 7, 9, 10 and 11;
neither neural networks nor LICORS appear to be optimal for any patient
across all forecast windows, although neural networks perform well for
patients 4, 5 and 8, while LICORS predicts well for patients 4, 8 and
11. Between-patient differences prevent any particular forecasting
method from dominating other methods across patients, but the LMAR
model seems to offer the most accurate overall point forecasts given
these results.

%Table~\ref{results12}, for the results of the 0.4~s forecast, tells a
%similar story, with ridge regression never outperforming both NNs and
%LMAR for any patient or error summary statistic, though comparable to
%NNs for patients 6, 8, 9, and 10. LMAR looks to be the most accurate
%for patients 6, 7, 9, 10, and 11, while neural networks do the best
%for patients 4, 5, and 8.

%The results in table~\ref{results18} (for the 0.6~s forecast window)
%are once again similar. Ridge regression is never optimal for any
%patient or error summary statistic, and the disparity is more evident
%in many of the patients than for shorter forecasting windows. LMAR
%does the best with this forecast window for patients 4, 6, 7, 10, and
%11. Neural netwoks do the best with patient 8, and patients 5 and 9
%have similar error summaries for both LMAR and NNs.

%Another important consideration of predictive performance is the
%proportion of forecasted observations for which one prediction method
%achieves a lower error than the other two methods. In other words,
%this is the proportion of time each predicted time series is closest
%to the true time series. Table~\ref{resultspred} gives these
%proportions for each patient/forecast window. We also provide the
%ratio of total absolute error to the minimum pointwise total absolute
%error among the three methods. Here again, we see LMAR performing very
%well for all forecast windows considered, particularly for patients 6,
%7, 9, 10, and 11.

%s5.3 #&#
\subsection{Qualitative summaries of point forecasts}\label{subsecqualsummaries}

\begin{figure}%[h]

\includegraphics{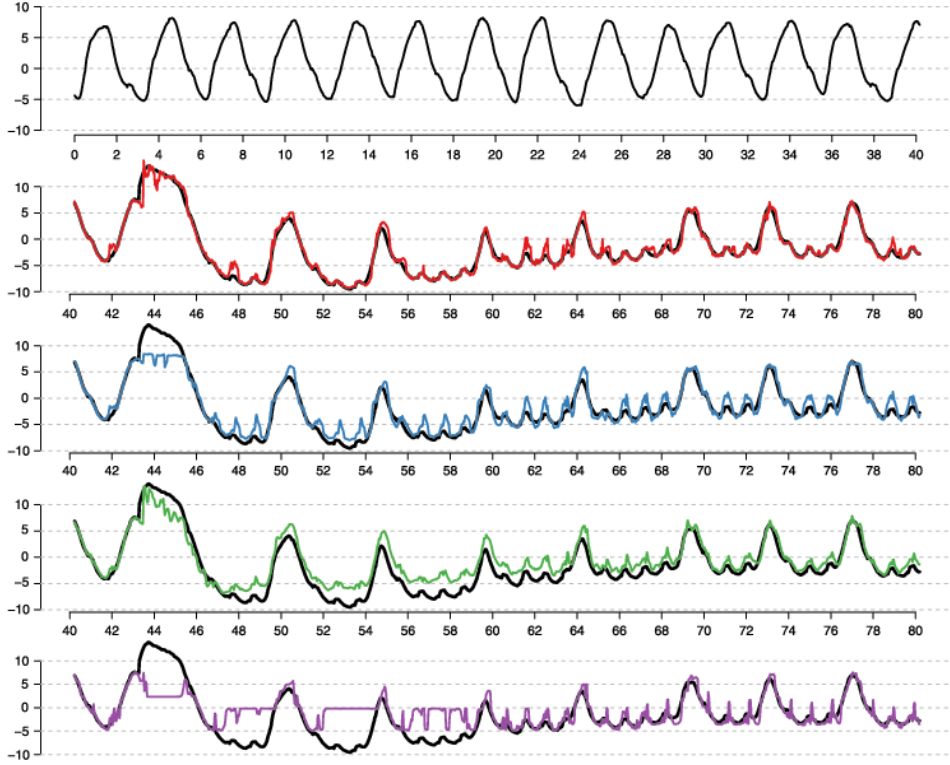}

\caption{Predictions for patient 9, day 3, beam 6 with a forecast
window of 0.2~s. Location (mm) is the $y$ axis and time (s) the $x$
axis. The 40~s training sequence is top, with predictions for the next
40~s from LMAR in red, NN in blue, ridge regression in green and LICORS
in purple.}
\label{pred205}
\end{figure}

When looking at the predicted time series for each method used, the
general pattern we observe is that LMAR outperforms the other three
methods when the data undergo changes in shape, period or
amplitude---or, more generally, when the test data do not resemble the
training data.
%, such as was highlighted in figure~\ref{motifs}.
Figure~\ref{pred205} shows one (atypically dramatic) instance of such
behavior. The top curve is the first 40 seconds of the time series, on
which all prediction methods were trained. The next four curves give
the predicted time series at a window of 0.2~s for LMAR (red), NN
(blue), ridge regression (green) and LICORS (purple). It is clear from
the figure that the end of the training period for this time series
coincided with a dramatic change in the patient's respiration.

%
%f7 #&#

Both neural networks and LICORS suffer from the range of the curve
being larger (dropping below $-$5~mm and exceeding 10~mm) after the
training period; for both methods, the training data bounds the range
of point forecasts, regardless of the input vector for future test
cases. For LICORS, when the test data is below the minimum of the
training data ($-$5~mm), the single predictive state associated with the
minimal values of the training data will dominate, leading to brief
periods of static forecasts. With this time series, this particular
predictive state represents an abrupt transition between sharp exhale
and sharp inhale. Thus, the forecasts for the test data are dramatic
overestimates throughout the ``U'' shaped motifs starting around
$t=47$, where the patient does not actually fully inhale.

Ridge regression seems to accurately predict the magnitudes of
increases and decreases, yet the predictions are off by a nearly
constant factor for $t\in(48, 68)$. In the context of the ridge
regression model \eqref{ridge}, this suggests that $\beta$ is correctly
specified, but perhaps $\beta_0$ is time varying. The LMAR model
includes an autoregressive term for the most recent $p$ observations in
its forecast, and thus, like ridge regression, accurately predicts
rates of change in the time series. Moreover, the stochastic
location-mixture component in the LMAR prediction adjusts predictions
for gradual magnitude shifts in the data.

%
%f8 #&#
\begin{figure}%[h]

\includegraphics{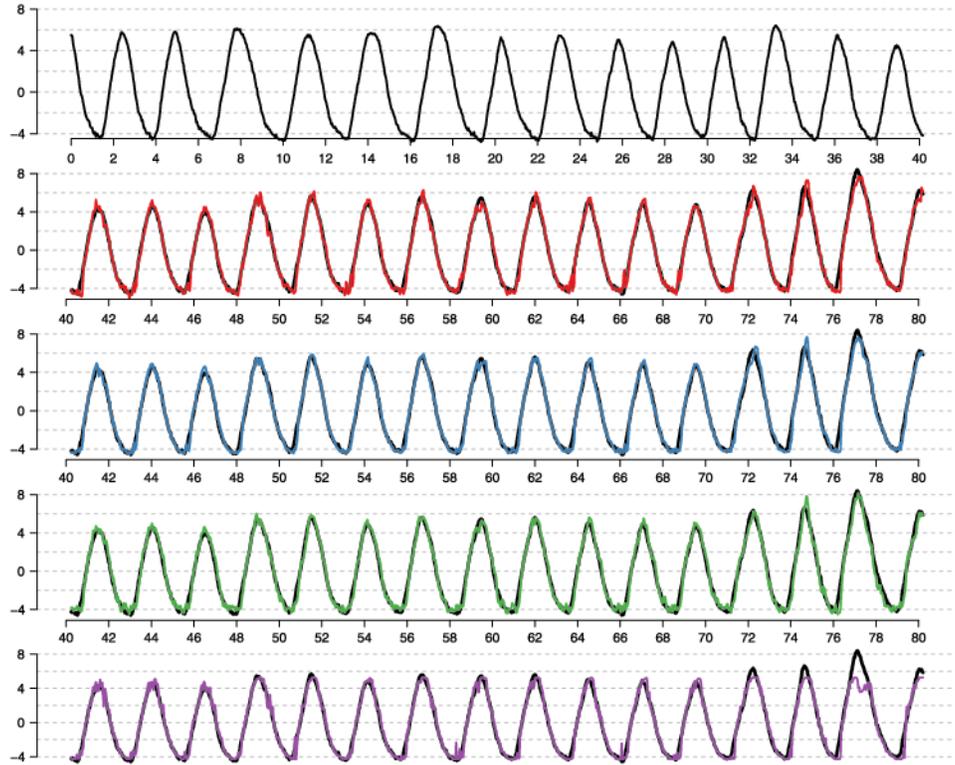}

\caption{Predictions for patient 4, day 6, beam 1 with a forecast
window of 0.2~s. Location (mm) is the $y$ axis and time (s) the $x$
axis. The 40~s training sequence is top, with predictions for the next
40~s from LMAR in red, NN in blue, ridge regression in green and LICORS
in purple.}
\label{pred124}
\end{figure}

Another reason why the LMAR model works relatively well when the test
data differ from the training data is that the form of the dependence
of forecasts on the most recent $p$ observations evolves, whereas it
remains static for the other three methods. While the parameters of the
model are not re-estimated during real-time prediction, LMAR uses the
entire history of the time series in making forecasts, not just the
first 40 seconds alongside the most recent $p$ observations, as is the
case with the other three methods. With appropriate parallel
computating resources, all methods could theoretically update
parameters continuously (or periodically) throughout treatment.
\citet{murphycomparative2006} continuously retrained neural
networks using the updated history of the respiratory trace. While they
did not compare this to the alternative of not actively updating the
forecast model, \citet{krausscomparative2011} did so and found a
small improvement in RMSE of about 1--3\%.

When the time series are more well behaved, all four methods perform
quite well; in fact, neural networks tend to have the lowest errors
when all four curves are accurate. Figure~\ref{pred124} shows the
training and prediction test series for a strongly periodic respiratory
trace. We should expect the performance of neural networks to be
superior when the dynamics of the tumor motion are stable, as the
parameter space for neural networks is far larger; in theory,
feedforward neural networks with at least one hidden layer can
approximate any continuous function arbitrarily well [\citet
{hornik1989multilayer}], including time series prediction.

%s5.4 #&#
\subsection{Interval and distributional forecasts}\label{subsecintervals}

Unlike commonly used time series models in the tumor-tracking
literature, the LMAR model provides multimodal, heteroskedastic
predictive distributions, which are theoretically appropriate for
forecasting respiratory motion. Despite this, our analysis of
predictive performance has focused exclusively on the accuracy of point
forecasts because in current implementations of tumor-tracking systems,
there is no clinical value in obtaining interval or distributional
forecasts. The treatment beam has a fixed width and is always on,
meaning an interval or distributional forecast does not alter the
optimal course of action of a tumor-tracking system already supplied
with a point forecast. However, interval/distributional forecasts would
prove valuable if we could, for instance, suspend the treatment beam
instantaneously if the predicted probability of the tumor location
being enclosed by the treatment beam fell below a certain threshold.

%
%t4 #&#
\begin{table}%[h!]
\tabcolsep=0pt
\caption{Summary of interval and distributional forecasts for all four
methods at all three forecast windows. The interval coverage considered
is $90\%$ confidence intervals. \emph{Log PS} refers to the log
probability score of the predictive distribution. For each metric, the
most desirable value among the four methods for each patient/forecast
window combination is in \textbf{bold}}\label{resultspred}
\begin{tabular*}{\tablewidth}{@{\extracolsep{\fill}}@{}lccccccc@{}}
\hline
& & \multicolumn{2}{c}{\textbf{0.2~s forecast}} & \multicolumn{2}{c}{\textbf{0.4~s forecast}} & \multicolumn{2}{c@{}}{\textbf{0.6~s forecast}}\\[-6pt]
& & \multicolumn{2}{c}{\hrulefill} & \multicolumn{2}{c}{\hrulefill} & \multicolumn{2}{c@{}}{\hrulefill}\\
\textbf{Patient} & \textbf{Method} & \textbf{Coverage} & \textbf{Log PS} & \textbf{Coverage} & \textbf{Log PS} & \textbf{Coverage} & \textbf{Log PS}\\
\hline
\phantom{0}{4} & LMAR & 0.84 & 0.72 & \textbf{0.86} & 1.30 & \textbf{0.93}& 1.37 \\
& NNs & 0.88 & \textbf{0.57} & 0.83 & 1.34 & 0.85 & 1.58 \\
& Ridge & 0.85 & 0.80 & 0.84 & 1.53 & 0.84 & 1.86 \\
& LICORS & \textbf{0.89} & 0.70 & 0.84 & \textbf{1.03} & 0.84 & \textbf{1.32}
\\[3pt]
\phantom{0}{5} & LMAR & \textbf{0.87} & \textbf{0.71} & \textbf{0.88} & \textbf{1.20} & \textbf{0.93} & \textbf{1.30} \\
& NNs & 0.85 & 0.72 & 0.78 & 1.52 & 0.80 & 1.75 \\
& Ridge & 0.85 & 0.91 & 0.84 & 1.53 & 0.82 & 1.91 \\
& LICORS & 0.84 & 1.04 & 0.82 & 1.46 & 0.79 & 1.78
\\[3pt]
\phantom{0}{6} & LMAR & \textbf{0.87} & 1.25 & \textbf{0.88} & \textbf{1.85} &\textbf{0.93} & \textbf{2.07} \\
& NNs & 0.79 & 1.31 & 0.74 & 2.16 & 0.76 & 2.53 \\
& Ridge & \textbf{0.87} & \textbf{1.22} & 0.85 & 1.91 & 0.83 & 2.26 \\
& LICORS & 0.79 & 1.58 & 0.70 & 2.57 & 0.66 & 2.82
\\[3pt]
\phantom{0}{7} & LMAR & 0.85 & \textbf{0.30} & \textbf{0.85} & \textbf{0.87} &\textbf{0.89} & \textbf{1.09} \\
& NNs & \textbf{0.88} & 0.48 & 0.84 & 1.35 & 0.84 & 1.82 \\
& Ridge & 0.86 & 0.63 & 0.83 & 1.49 & 0.82 & 1.95 \\
& LICORS & 0.84 & 0.78 & 0.77 & 1.16 & 0.76 & 1.59
\\[3pt]
\phantom{0}{8} & LMAR & \textbf{0.89} & 1.67 & 0.91 & 2.30 & 0.94 &2.60 \\
& NNs & 0.94 & \textbf{1.53} & 0.82 & 2.36 & \textbf{0.90} & 2.59 \\
& Ridge & 0.88 & 1.82 & 0.85 & 2.51 & 0.82 & 2.90 \\
& LICORS & 0.94 & 1.71 & \textbf{0.90} & \textbf{2.11} & 0.88 & \textbf{2.39}
\\[3pt]
\phantom{0}{9} & LMAR & \textbf{0.89} & \textbf{0.87} & \textbf{0.90} & \textbf{1.65} & \textbf{0.92} & \textbf{2.07} \\
& NNs & 0.86 & 1.02 & 0.78 & 2.20 & 0.80 & 2.77 \\
& Ridge & 0.81 & 1.54 & 0.81 & 2.21 & 0.81 & 2.64 \\
& LICORS & 0.86 & 1.62 & 0.81 & 1.98 & 0.79 & 2.31
\\[3pt]
{10} & LMAR & \textbf{0.86} & \textbf{1.18} & \textbf{0.88} & \textbf{1.94} & \textbf{0.91} & 2.33 \\
& NNs & 0.84 & 1.23 & 0.76 & 2.25 & 0.79 & 2.65 \\
& Ridge & 0.83 & 1.35 & 0.84 & 2.03 & 0.84 & 2.44 \\
& LICORS & \textbf{0.86} & 1.61 & 0.82 & 2.02 & 0.81 & \textbf{2.31}
\\[3pt]
{11} & LMAR & 0.85 & \textbf{1.38} & \textbf{0.87} & 2.13 & \textbf{0.91} & 2.36 \\
& NNs & 0.87 & 1.50 & 0.80 & 2.70 & 0.83 & 2.91 \\
& Ridge & 0.86 & 1.63 & 0.85 & 2.44 & 0.85 & 2.84 \\
& LICORS & \textbf{0.88} & 1.56 & 0.83 & \textbf{1.99} & 0.82 & \textbf{2.25} \\
\hline
\end{tabular*}\vspace*{-3pt}
\end{table}

Table~\ref{resultspred} gives a summary of the performance of
out-of-sample interval and distributional forecasts to complement the
summaries of point forecasts. The LMAR model, by specifying a
data-generating process, naturally provides full predictive
distributions as a by-product of point prediction. The same is true for
ridge regression (assuming the typical homoskedastic Gaussian structure
for the residuals) and LICORS. Neural networks do not naturally provide
predictive distributions; following \citet
{tibshirani1996comparison}, we obtain them by bootstrapping, while
assuming prediction errors are (heteroskedastic) independent Gaussians,
with mean 0 and variance estimated by bootstrapping.

We expect LMAR prediction intervals to undercover, since uncertainty in
the estimation of $\Sigma$ is omitted from our forecasts. While this is
indeed the case, for all patients and forecast windows, $90\%$
prediction intervals have between $84\%$ and $94\%$ coverage---a more
appropriate range than any other method can claim.

The logarithmic score in Table~\ref{resultspred} refers to the negative
logarithm of the predictive density evaluated at the true observation,
averaged over each out-of-sample prediction (the result in Table~\ref{resultspred} then averages each of these scores over all beams from
the same patient). The logarithmic score is a \textit{proper} scoring
rule---its expected value is minimized by the oracle (or true)
predictive distribution---thus, lower values indicate a better fit
between the predictive distributions and realized values of a patient's
time series [\citet{gneiting2007probabilistic}].

Generalizing across patients and forecast windows, in comparison to the
other methods considered, the LMAR model seems to most accurately
characterize prediction uncertainty.

%s6 #&#
\section{Discussion}\label{secdiscussion}
The location-mixture autoregressive (LMAR) model introduced in this
paper provides accurate, real-time forecasts of lung tumor motion. Our
method achieves better performance on out-of-sample prediction for
forecasts windows of 0.2~s, 0.4~s and 0.6~s for the majority of the
patients considered than existing methods such as neural networks
[which performed best in a prediction comparison study of \citet
{krausscomparative2011}] and penalized linear models (a common baseline
for judging predictive performance). We also note that uncertainty
quantification is quite straightforward using our model, whereas it is
hard to do using neural networks.

The LMAR model is similar to other autoregressive models that yield
multimodal conditional distributions, such as the class of threshold
autoregressive models [\citet{tong1978threshold}], yet the
parameter space consists of just a single, low-dimensional covariance
matrix, and the model admits accurate closed-form\vadjust{\goodbreak} approximations of
multiple-step ahead predictive distributions. The LMAR model also has a
useful interpretation in the context of time series motifs, which can
\mbox{describe} the data-generating process and the form of forecasts.

While the predictive performance of our method on this data set is very
encouraging, the parameter inference for the LMAR model presented here
is approximate, and the assumptions of both the model and its inference
may not be appropriate for some other nonlinear time series.
Formalizing and generalizing the LMAR model is thus a fruitful area for
future work.

Real-time prediction of lung tumor motion presents additional
challenges to those presented in this work. It is preferable to have as
short a training window as possible, since during this time the patient
may be irradiated without actually receiving the benefit of tumor
tracking. While some training is actually necessary to estimate the
system latency in some cases (we have treated it as fixed throughout
this work), the 40 seconds used for training in this paper (while
typical in the literature on the subject) could ideally be reduced.

Also, one can consider patient-specific hyperparameter values and/or
tuning parameters or modify the model to borrow information across the
patients. Due to the need for real-time model fitting before we can
forecast, it is most likely infeasible to apply any model selection
criteria (either within-model, such as for hyperparameters, or
between-model) after having begun to observe data. More study of
between-patient and within-patient variability in model fits could help
researchers use more patient-optimal prediction methods (as well as
begin prediction after a shorter training sequence, as they would not
need to rely solely on the observed data for parameter estimation).

The parametric simplicity of the LMAR model, as well as its
formalization as a statistical model as opposed to a prediction
algorithm, enable generalizations of our procedure to include
hierarchical models and other statistical structures that address the
challenges of delivering accurate external beam radiotherapy. Combined
with its excellent predictive performance on real data, the LMAR model
represents a promising new contribution to this area of research.

\section*{Acknowledgments}
The authors would like to thank Dr. Seiko Nishioka of the Department of Radiology, NTT Hospital,
Sapporo, Japan, and Dr. Hiroki Shirato of the Department of Radiation Medicine, Hokkaido University
School of Medicine, Sapporo, Japan, for sharing the patient tumor
motion data set with us.
The content is solely the responsibility of the authors and does not necessarily represent the
official views of the National Cancer Institute, National Science Foundation or the National Institutes of Health.

\begin{supplement}[id=suppB]
\sname{Supplement}
\stitle{Code}
\slink[doi]{10.1214/14-AOAS744SUPP} %[doi] - jei reikia suskaldyti doi
\sdatatype{.zip}
\sfilename{aoas744\_supp.zip}
\sdescription{\texttt{R} Code used for fitting and forecasting with
the LMAR model.}
\end{supplement}

% imsref loaded by linak, 2014-05-14 16:06:28
% imsref loaded by linak, 2014-05-14 16:13:23
%
% imsref loaded by linak, 2014-05-16 14:36:19

\printaddresses
\end{document}